\PassOptionsToPackage{table,xcdraw}{xcolor}

\documentclass[format=sigconf, review=false, anonymous=false, screen=true, dvipsnames]{acmart}

\copyrightyear{2026}
\acmYear{2026}
\setcopyright{cc}
\setcctype{by}
\acmConference[CHI '26]{Proceedings of the 2026 CHI Conference on Human Factors in Computing Systems}{April 13--17, 2026}{Barcelona, Spain}
\acmBooktitle{Proceedings of the 2026 CHI Conference on Human Factors in Computing Systems (CHI '26), April 13--17, 2026, Barcelona, Spain}
\acmPrice{}
\acmDOI{10.1145/3772318.3790329}
\acmISBN{979-8-4007-2278-3/2026/04}

\usepackage{booktabs} 

\usepackage{enumitem}
\setlist[itemize]{leftmargin=*}
\setlist[enumerate]{leftmargin=*,label=\arabic*.}

\usepackage{xcolor}

\usepackage{xspace}

\usepackage{graphicx}

\usepackage{verbatim}

\AtBeginDocument{%
 \abovedisplayskip=3pt plus 3pt minus 3pt
 \abovedisplayshortskip=3pt plus 3pt minus 3pt
 \belowdisplayskip=3pt plus 3pt minus 3pt
 \belowdisplayshortskip=3pt plus 3pt minus 3pt
}

\makeatletter
\newcommand{\pquote}[1]{%
  \@ifnextchar\bgroup
    {\pquote@with{#1}}%
    {\pquote@without{#1}}%
}

\newcommand{\pquote@without}[1]{%
  \textit{``#1''}%
}

\newcommand{\pquote@with}[2]{%
  \textit{``#1''}{\,}{\small[#2]}%
}
\makeatother

\makeatletter
\newcommand{\pquotemid}[1]{%
  \@ifnextchar\bgroup
    {\pquotemid@with{#1}}%
    {\pquotemid@without{#1}}%
}

\newcommand{\pquotemid@without}[1]{%
  \begingroup
    \leftskip1cm\relax
    \rightskip1cm\relax
    \textit{``#1''}%
  \endgroup
}

\newcommand{\pquotemid@with}[2]{%
  \begingroup
    \leftskip1cm\relax
    \rightskip1cm\relax
    \textit{``#1''}{\,}{\small[#2]}%
  \endgroup
}
\makeatother

\newcommand{\ie}{\emph{i.e.,{\xspace}}}
\newcommand{\eg}{\emph{e.g.,{\xspace}}}
\newcommand{\vs}{vs.{\xspace}}

\newcommand{\f}[1]{{\textsc{#1}}\xspace}

\newcommand{\m}[1]{{\textit{#1}}\xspace}

\newcommand{\by}{$\times$\xspace}

\newcommand{\mean}[1]{$\textsc{m}\!=\!#1$}
\newcommand{\sd}[1]{$\textsc{sd}\!=\!#1$}

\makeatletter
\newcommand{\p}[1]{%
  \@ifnextchar\bgroup
    {\p@with{#1}}%
    {\p@without{#1}}%
}

\newcommand{\p@without}[1]{%
  {\small\mbox{$p<#1$}}%
}

\newcommand{\p@with}[2]{%
  {\small\mbox{$p#2#1$}}%
}
\makeatother

\makeatletter
\newcommand{\rvalue}[1]{%
  \@ifnextchar\bgroup
    {\rvalue@with{#1}}%
    {\rvalue@without{#1}}%
}

\newcommand{\rvalue@without}[1]{%
  {\small\mbox{$r<#1$}}%
}

\newcommand{\rvalue@with}[2]{%
  {\small\mbox{$r#2#1$}}%
}
\makeatother

\makeatletter
\newcommand{\dvalue}[1]{%
  \@ifnextchar\bgroup
    {\dvalue@with{#1}}%
    {\dvalue@without{#1}}%
}

\newcommand{\dvalue@without}[1]{%
  {\small\mbox{$d=#1$}}%
}

\newcommand{\dvalue@with}[2]{%
  {\small\mbox{$r#2#1$}}%
}
\makeatother

\makeatletter
\newcommand{\anova}[4]{%
  \@ifnextchar\bgroup
    {\anova@with{#1}{#2}{#3}{#4}}%
    {\anova@without{#1}{#2}{#3}{#4}}%
}

\newcommand{\anova@without}[4]{%
  {\small$F_{#1,#2}=#3$, $p<#4$}%
}

\newcommand{\anova@with}[5]{%
  {\small$F_{#1,#2}=#3$, $p<#4$, $\eta_G^2=#5$}%
}
\makeatother

\makeatletter
\newcommand{\rmanova}[4]{%
  \@ifnextchar\bgroup
    {\rmanova@with{#1}{#2}{#3}{#4}}%
    {\rmanova@without{#1}{#2}{#3}{#4}}%
}

\newcommand{\rmanova@without}[4]{%
  {\small$F_{#1,#2}=#3$, $p<#4$}%
}

\newcommand{\rmanova@with}[5]{%
  {\small$F_{#1,#2}=#3$, $p<#4$, $\eta_P^2=#5$}%
}
\makeatother

\definecolor{GREEN}{rgb}{0.0,0.6,0.0}
\definecolor{BLUE}{rgb}{0.0,0.2,0.7}
\definecolor{GOLD}{rgb}{0.6,0.6,0.0}
\definecolor{CYAN}{rgb}{0.0,0.5,0.5}
\definecolor{PURPLE}{rgb}{0.5,0.0,0.5}

\definecolor{RED}{rgb}{0.7,0.0,0.0}
\definecolor{GRAY}{gray}{0.5}

\definecolor{LIGHTGRAY}{HTML}{DDDDDD}

\definecolor{GOLD}{HTML}{FFE052}

\makeatletter
\g@addto@macro\normalsize{%
  \setlength\abovedisplayshortskip{-9pt}
  \setlength\belowdisplayshortskip{3pt}
}
\makeatother

\newcommand{\rev}[1]{#1}

\definecolor{REVISIONRED}{HTML}{9e092f}
\definecolor{REVISIONGREEN}{HTML}{427a11}
\definecolor{REVISIONGOLD}{HTML}{806b06}

\usepackage{booktabs}
\usepackage{hyperref}
\hypersetup{
    colorlinks=true,
    linkcolor=blue,
    filecolor=magenta,      
    urlcolor=cyan,
}

\newcommand{\UP}{Uncertain Pointer\xspace}

\makeatletter
\g@addto@macro\normalsize{%
  \setlength\abovedisplayshortskip{-9pt}
  \setlength\belowdisplayshortskip{3pt}
}
\makeatother

\newcommand{\EX}{%
  \raisebox{-0.2em}{\includegraphics[height=1.3em]{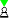}}~\f{external}%
}

\newcommand{\IN}{%
  \raisebox{-0.2em}{\includegraphics[height=1.3em]{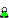}}~\f{internal}%
}

\newcommand{\BOUND}{%
  \raisebox{-0.2em}{\includegraphics[height=1.3em]{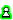}}~\f{boundary}%
}

\newcommand{\FILL}{%
  \raisebox{-0.2em}{\includegraphics[height=1.3em]{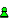}}~\f{fill}%
} 

\begin{document}

\tolerance=400 

\title[Uncertain Pointer]{Uncertain Pointer: Situated Feedforward Visualizations for Ambiguity-Aware AR Target Selection}




\author{Ching-Yi Tsai}
\orcid{0000-0001-5664-6562}
\email{ching-yi@princeton.edu}
\affiliation{%
  \institution{Princeton University}
  \city{Princeton}
  \state{NJ}
  \country{USA}
}

\author{Nicole Tacconi}
\orcid{0009-0005-2417-2899}
\email{nicole.tacconi@princeton.edu}
\affiliation{%
  \institution{Princeton University}
  \city{Princeton}
  \state{NJ}
  \country{USA}
}

\author{Andrew D. Wilson}
\orcid{0000-0001-5751-9354}
\email{awilson@microsoft.com}
\affiliation{%
  \institution{Microsoft Research}
  \city{Redmond}
  \state{WA}
  \country{USA}
}

\author{Parastoo Abtahi}
\orcid{0009-0000-2145-3445}
\email{parastoo@princeton.edu}
\affiliation{%
  \institution{Princeton University}
  \city{Princeton}
  \state{NJ}
  \country{USA}
}

\renewcommand{\shortauthors}{Tsai et. al.}

\begin{abstract}
Target disambiguation is crucial in resolving input ambiguity in augmented reality (AR), especially for queries over distant objects or cluttered scenes on the go. Yet, visual feedforward techniques that support this process remain underexplored. We present Uncertain Pointer, \rev{a} systematic exploration of feedforward visualizations that annotate multiple candidate targets before user confirmation, either by adding distinct visual identities (e.g., colors) to support disambiguation or by modulating visual intensity (e.g., opacity) to convey system uncertainty. First, we construct a pointer space of 25 pointers by analyzing {existing} placement strategies and visual signifiers used in target visualizations across 30 years of relevant literature. We then evaluate them through two online experiments (n = 60 and 40), measuring user preference, confidence, mental ease, target visibility, and identifiability across varying object distances and sparsities. Finally, from the results, we derive design recommendations in choosing different Uncertain Pointers based on AR context and disambiguation techniques.


\end{abstract}

%
%



\begin{teaserfigure}
  \includegraphics[width=\textwidth]{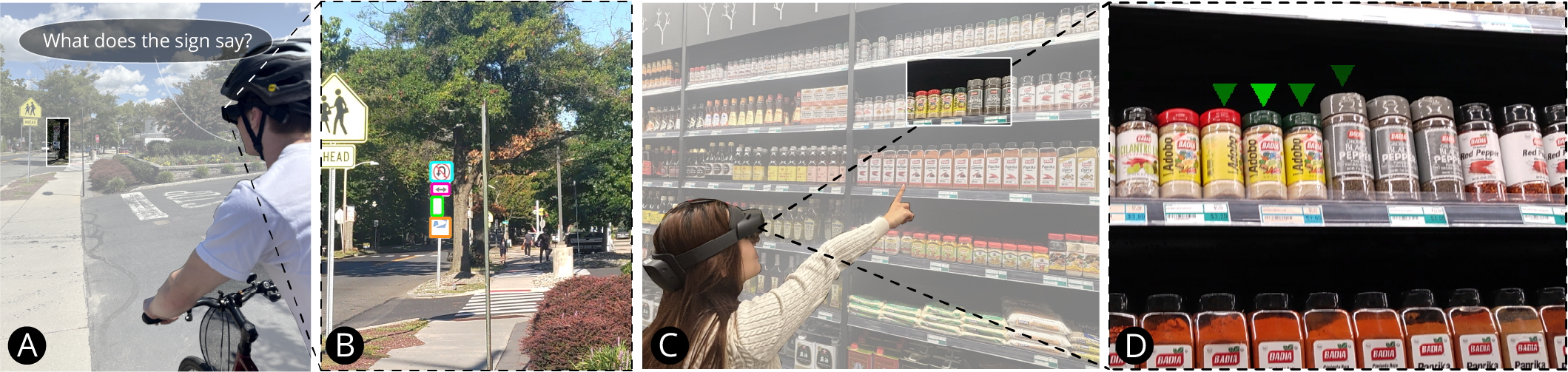}
  \caption{Uncertain Pointer explores feedforward visualizations that convey input ambiguity and facilitate disambiguation. For example, when (a) speech input is ambiguous, (b) visual identifiers such as color labels can distinguish candidates and support verbal clarification (\eg{} saying ``the pink one''), or when (c) users' pointing input to query grocery items refers to multiple possible targets, (d) Uncertain Pointers help narrow the candidate set to aid selection (\eg{} using reduced control-display gain) while keeping the system’s interpretation transparent.}
  \Description{Figure 1 is a four-panel montage illustrating how the Uncertain Pointer system conveys ambiguity and supports disambiguation in AR.
Panel (a) shows a real-world street scene where multiple traffic signs appear in the distance. The system overlays identical visual markers on several signs, indicating that the user's verbal query (“that sign”) is linguistically ambiguous.
Panel (b) adds distinct color-coded labels (e.g., pink, yellow, blue) above each candidate sign. These identifiers provide disambiguating attributes that allow spoken clarification such as “the pink one.” The scene emphasizes how identity-based pointers support multimodal refinement.
Panel (c) depicts a grocery aisle where the user points their hand toward a shelf of items. Multiple objects along the pointing ray are highlighted simultaneously, showing that gestural input can map to multiple plausible targets when depth or alignment is ambiguous.
Panel (d) demonstrates a level-based pointer: targets are annotated with graded intensity cues (e.g., varying opacity, size, or saturation) to reflect differential system confidence. Higher-likelihood targets appear more salient, and lower-likelihood ones fade subtly. This preview helps the user understand uncertainty and adjust their pointing or gaze to resolve ambiguity.
Overall, the figure illustrates how Uncertain Pointer visually represents candidate target sets, highlights the system's interpretation, and offers mechanisms for refining ambiguous input.}
  \label{fig:teaser}
\end{teaserfigure}

\maketitle



\section{Introduction}

Pointing or referencing an object or region of interest is a fundamental part of human action, helping to establish shared attention and ground the subject of interaction ~\cite{clark2003pointing, clark1991grounding}.
Recent advances in display technology, spatial tracking, and AI have enabled this behavior to extend into augmented reality (AR). Smart glasses and AR headsets (\eg{} Meta Orion\footnote{Meta Orion \url{https://about.meta.com/realitylabs/orion/}}, Snapchat Spectacles\footnote{Snapchat Spectacles \url{https://www.spectacles.com/}}, and VIVE Eagle\footnote{VIVE Eagle \url{https://www.vive.com/us/product/vive-eagle/overview/}}) now support multimodal interactions that allow users to point, gesture, and speak about real-world objects in situ, anytime, and on the go.

However, despite their ubiquity and convenience, such queries are often prone to ambiguity in user input.
For example, although voice input is natural ~\cite{liszkowski2012prelinguistic, diessel2020demonstratives}, verbal commands, in particular, are prone to linguistic ambiguity (\eg{} syntactic ambiguity~\cite{taha1983types}  or pronoun ambiguity), a long-standing challenge dating back to Bolt's visionary ``Put That There'' system~\cite{putthatthere} and its successors~\cite{cohen_quickset_1997, koons1993integrating, zhai_magic_1999}.
Moreover, input on-the-go introduces tradeoffs in stability and precision, including requiring significantly more interaction time ~\cite{zhou_intonthego_2016} and strategies to compensate for input noise~\cite{kane_walkingpublic_2008}.
Lastly, AR-based target queries can be affected by target complexity, such as object distance~\cite{Matt_distanttarget_2018} and  clutter~\cite{jirak_cluttered_2021}, which hinder both system recognition and user specification.
To illustrate, consider a wearer of smart glasses passing down the street (Figure~\ref{fig:teaser}a), gazing toward an upcoming group of traffic signs and asking, ``What does that sign say?'' The system must resolve multiple ambiguities: ambiguous linguistic reference (``that sign''), noisy gaze-pointing due to motion, and multiple plausible targets being far and densely packed.

To address such input ambiguity,  
researchers have investigated {explicit} disambiguation techniques that require users to take additional steps to refine their inaccurate input and disambiguate one target from many ~\cite{kopper_progressiverefinement_2011, grossman_3Dvolumeselect2006, yu_fullyoccluded_2020, sadana_expandingselection_2016, wu_horizontaldragger_2015}.
Within these disambiguation methods, visualizations are commonly integrated as a lasso, a colored area, or a bounded region to indicate an initial coarse selection. However, they are seldom evaluated or systematically compared. Furthermore, most existing visualizations for disambiguation have been developed and tested in plain scenes, where object characteristics are tightly controlled. In contrast, real-world objects in AR scenarios vary significantly in distance and density. As a result, determining which visualizations are best suited for different AR disambiguation contexts remains an open question.

In this paper, we move beyond input to {investigate} \textit{visualizations} that support disambiguation and convey uncertainty during AR target selection across varied object layouts. 
Our key observation is that {explicit} disambiguation techniques begin with an ambiguous, coarse selection step, in which a set of candidate objects is identified. To support this, we explore visualizations that annotate multiple potential targets in AR, which we term \UP{}s.\footnote{\rev{Terminology: Throughout the paper, we use \UP (singular) to denote our work as a whole, and \UP{}s (plural) to denote the set or any subset of specific pointer designs and candidates.}} Specifically, we investigate three types of pointer designs:
(1) \textbf{Level pointers} (Figure~\ref{fig:teaser}c \& d) modulate visual intensity of annotations (\eg{} size, opacity) to convey system uncertainty or selection likelihood transparently. The graded information also inherently enables spatial disambiguation via gaze or pointing as there is a clear indication of the direction in which the user shall adjust their input.
(2) \textbf{Identity pointers} (Figure~\ref{fig:teaser}a \& b) assign distinct, non-hierarchical identities (\eg{} color, letters) to each candidate without implying likelihood, supporting verbal disambiguation via added unique and descriptive attributes. (3) \textbf{Certain pointers} annotate a single object, highlighting the system’s single top selection directly. We include this as a baseline, but also for a highly certain AR selection scenario and ~{implicit} disambiguation techniques that do not include a first-step coarse selection.
While each pointer type is suited for different disambiguation strategies, they are not mutually exclusive. Systems can combine them, \eg{} using Level Pointers for first-step coarse selection and then Identity or Certain Pointers for finer verbal resolution within the selected targets set. Thus, our investigation offers flexible visualization choices adaptable to various disambiguation workflows, along with their tradeoffs and design considerations.
To explore potential designs for \UP{}s, we first conducted a holistic systematic survey into target-dependent visualization techniques regarding feedforward visualization, uncertainty visualization, and situated visualization from the past 30 years of publications at ACM CHI, UIST, DIS, VRST, SUI, and AutomotiveUI, along with IEEE {TVCG, ISMAR,} 3DUI, and VR. This survey informed the characterization of the four major pointer archetypes (\ie{} \f{external}, \f{internal}, \f{boundary}, and \f{fill}) along with the four most prevalent visual signifiers (\ie{} \f{color}, \f{size}, \f{opacity}, and \f{text}), which can be combined with our three pointer types that convey different degrees of uncertainty complexity (\ie{} \f{certain}, \f{identity}, and \f{level}). We explore how these archetypes, signifiers, and uncertainty complexities can be combined to form new possible visualizations, resulting in \UP's pointer space of 25 candidates.

We then conducted two pre-registered, online user studies to evaluate how well \UP{}s communicates system uncertainty, improves target noticeability, and
minimizes occlusion, along with subjective metrics such as user preference and perceived mental effort. The evaluations covered 4 target-complexity scenarios (near/far × dense/sparse) and 3 target-count levels. To manage the number of conditions per study, Study 1 (n = 60) focused on \f{certain} and \f{identity} Pointers, which informed the exclusion of low-performing designs. Study 2 (n = 40) then evaluated \f{level} visualizations. Finally, based on the findings, we derived design recommendations and usage examples for applying \UP{}s across AR scenarios. 

In sum, this work systematically investigates \UP{}s to communicate input uncertainty for AR target selection:
\begin{itemize}[leftmargin=*]
    \item  We conducted a literature survey to categorize existing feedforward and uncertainty visualizations (at {TVCG}, CHI, {ISMAR}, DIS, UIST, IEEE VR, 3DUI, AutomotiveUI, SUI, and VRST conferences) and generate \UP's pointer space.
    \item We investigated the effectiveness of \UP{}s and their trade-offs using two preregistered online studies (n = 60 and 40, respectively) across different background and target complexities.
    \item We provided design recommendations and example uses for future systems utilizing \UP{}s.
\end{itemize}


\section{Related Work}

\subsection{Selection Ambiguity and HCI}
Since the introduction of Fitts’ Law to HCI~\cite{mackenzie2018fitts}, it has been clear that small and distant targets are harder to select. As computing form factors evolved, more sources of input ambiguity emerged, including motor limitations (\eg{} the fat-finger problem), input during movement~\cite{zhou_intonthego_2016}, linguistic ambiguity~\cite{liszkowski2012prelinguistic, diessel2020demonstratives}, and system recognition errors. To address these, researchers have explored \textit{implicit disambiguation} that ~{utilizes additional input information or performs extra analysis on input behavior}~\cite{jae_gazepointar_2024, haan_intenselect_2005, schmidt_frustumpointing_2006, Steinicke2006, su_implicitfan_2014, ikeno_stickycursor_2023} and \textit{explicit disambiguation} strategies that require users to perform refinement or clarification for final selection~\cite{baloup2019raycursor, grossman_3Dvolumeselect2006, kopper_progressiverefinement_2011, yu_fullyoccluded_2020}.

{
For implicit disambiguation in 2D interfaces, Bubble Cursor~\cite{grossman_bubblecursor_2005} utilizes target proximity to dynamically resize its activation area and acquire the closest target. MAGIC mouse technique leverages gaze information to improve clicking accuracy ~\cite{fares_mousemagic_2013}. 
Other systems use icon semantics~\cite{chai_probresolution_2004}, voice~\cite{oviatt_mutualdisambiguation_1999}
, or statistical criteria from users' pointing ~\cite{li_bayesianpointer_2018, schmidt_frustumpointing_2006} 
and touch input~\cite{schwarz_framework_2010, xi_bayesiantouch_2013} behaviors.
For VR and AR, some implicit disambiguation techniques extend existing 2D techniques to 3D (\eg{} 3D bubble cursor ~\cite{vanacken_3Dbubblecursoranddepthraycursor} or selection-by-volume
~\cite{haan_intenselect_2005}), while others adopt depth estimation~\cite{mardanbegi_ambiguityVORdepth_2019}, 3D gestural recognition ~\cite{kaiser_mutualdisambiguation_2003, ma_focalpoint_2023}, or mobile gaze prediction ~\cite{barz_erroraware_2018}, along with techniques that leverage the human's inherent multimodal habit in 3D spatial interaction, such as utilizing gaze attention during speech query~\cite{jae_gazepointar_2024} and speech with gestural input ~\cite{lee_usabilitymultimodal_2013, olwal_senseshapes_2003}. 
}

Despite their unobtrusiveness, implicit disambiguation methods fail \rev{under degraded sensing conditions or when user intent cannot be reliably inferred. For example, when the sensing channel is noisy, critical cues are occluded or unavailable, or the user input is ambiguous or incomplete for the intended input channel. These issues are especially common for lightweight, ubiquitous devices in uncertain, dynamic contexts, where occlusion, motion jitter, and missing data are routine}.
 In contrast, explicit disambiguation requires users to take an additional step to resolve input ambiguity before final confirmation, helping bring the uncertainty to the user's awareness and prevent incorrect selections. 
Explicit approaches frequently incorporate (1) visualizations to highlight the initial coarse selection and (2) an input method for refinement and disambiguation. In terms of refinement modality, hand input is the most common one, including controller input to refine raycursor or pointing selection~\cite{vanacken_3Dbubblecursoranddepthraycursor, baloup_raycursor_2019, feiner2003flexible}, pointing to compensate initial region selection~\cite{wu_horizontaldragger_2015}, gestural input to specify target from multiple options within a 3D volume~\cite{grossman_multifingervolumetric_2004, grossman_3Dvolumeselect2006}, cluster of objects~\cite{yu_fullyoccluded_2020, kopper_progressiverefinement_2011}, or a list of potential operations ~\cite{chen_disambiguationobjectmanipulVR_2020}. Additionally, explicit techniques may utilize different input modalities from the initial input for confirmation or disambiguation, such as using head movement with gaze input~\cite{kyto_pinpointing_2018, kurauchi_hmagic_2015, spakov_lookandlean_2014, spakov_imsplegazehead_2012}, hand input with gaze refinement~\cite{kyto_pinpointing_2018,zhai_cascadedMAGIC_1999}, gaze cursor with hand adjustment~\cite{pfeuffer_gazeandtouchtablets_2016, chatterjee_gazeplusgesture_2015,zhang_eyecursorhand_2008,stellmach_lookandtouch_2012}, or adaptively switching fall-back modalities based on tasks and use cases~\cite{sidenmark_weightedpointer_2022}.
Regarding visualization, techniques such as color underlines and box outlines~\cite{sadana_expandingselection_2016}, circled areas~\cite{wu_horizontaldragger_2015, kopper_progressiverefinement_2011}, and cones or lassos~\cite{yu_fullyoccluded_2020} are often used when there are several potential candidate targets.

In summary, there is vast research on disambiguation techniques. {\UP{}'s exploration aims to complement disambiguation over targets in AR by exploring pointer design possibilities and their trade-offs. Specifically, our investigation on Certain pointer provides recommendations for implicit disambiguation, while our exploration on Identity and Level design assists disambiguation with object pointing~\cite{guiard_objectpointing_2004} in AR and verbal clarification over targets during interaction with visual assistants~\cite{wang2025resolving}.}

\subsection{Feedforward Visualization}
Feedforward refers to cues that indicate the possible outcomes of an action before it is executed, helping users anticipate system behavior and direct themselves to their goal~\cite{bridgenorman,norman2013design, interactionfrogger}. 
In desktop interfaces, feedforward is prevalent: hover effects highlight interactive elements, cursor position previews where interaction will occur, and cursor icon changes (\eg{} arrows to hand) indicate available actions. Beyond these common cues, researchers have explored rich visual feedforward strategies. For instance, ~\citet{guillon_targetexpansion_2015} introduces feedforward for target expansion, which dynamically enlarges a target’s effective area as the cursor approaches to enhance target recognition during selection. Fortunettes ~\cite{fortunettes} extend feedforward to widgets by previewing their future states (\eg{} previewing a checked checkbox to see on-click UI state changes). 
On a touch interface, ShadowGuides ~\cite{shadowguides} use projected visual elements such as dynamic arrows and keyframes to guide gesture before users complete their input, thus reducing learning burden.  

The most closely related work to ours involves feedforwards that visualize multiple input possibilities for user confirmation. For example, OctoPocus~\cite{bau_octopocus_2008} provides real-time visual previews of stroke-based gestures, using color and opacity to distinguish between multiple candidate gestures. Later adaptations extended this technique: ~\citet{malloch_fieldpathward_2017} incorporated gradient encodings to signal gesture likelihood (including improbable paths) and ~\citet{delamare_3doctopocus_2016} ports it to 3D input settings.
Probabilistic frameworks have been proposed to support such adaptive feedforwards for previewing multiple system actions~\cite{mankoff_abiguityresolution_2007, mankoff_toolkit_2000, schwarz_framework_2010}. For example, Schwarz et al.~\cite{schwarz_probframework_2015} introduces a probabilistic framework in which feedforward cues adapt dynamically: alternative actions (\eg{} play vs. add to playlist) are previewed, with the likely option emphasized to reveal the system's interpretation and guide user choice.

Together, these techniques demonstrate how feedforward can surface input uncertainty and guide resolution during interaction. However, they are mostly designed for 2D touchscreens and desktops, where input is bounded by standardized UI elements. \UP explores a similar idea, showing users multiple possible selection targets before they finalize a choice, but specific for real-world targets in AR, which requires new design considerations for situated, spatially anchored feedforward visualizations.


\subsection{Uncertainty Visualization for Interactive System}
Uncertainty visualization uses visual encodings to represent incomplete information or confidence levels, enabling users to better interpret and act on uncertain data ~\cite{skeels2010revealing}. Common techniques include error bars in bar charts, uncertainty ranges in line graphs, and color variations in heatmaps. These approaches typically map uncertainty to visual channels such as color, size, or opacity, complementing depictions of the single average value or state ~\cite{padilla2020uncertainty, tamara_2015_vizAD}.

In 2D interactive systems, uncertainty visualizations have supported user understanding across diverse domains, such as visualizing confidence in geospatial data~\cite{kong_audiovisual_2019}, hurricane forecasts~\cite{bica_hurricane_2019}, fertility predictions~\cite{schneider_fertility_2019}, GPS accuracy~\cite{ranasinghe_gps_2019}, and bus arrival times~\cite{kay_bus_2016}.
In 3D or physically situated environments, researchers have also applied similar principles to support decision-making. For example, uncertainty has been visualized through AR head-up displays to convey driving confidence~\cite{kunze_ardriveUV_2018}, or via in-car light bars to communicate road guidance uncertainty~\cite{kunze_drivingsafety_2017}. Others use spatial overlays to visualize sensor confidence around sensing devices~\cite{kim_sensorviz_2022}.

\UP{} draws on these works by bridging uncertainty visualization and AR pointer design. We explore different visual encodings, such as size and color, to communicate system confidence in candidate targets, supporting both transparency and user disambiguation. In particular, we focus on visualization techniques that emphasize targets by adding controlled levels of visual saliency. An alternative approach common in uncertainty visualization is to de-emphasize less certain referents (\eg{} through blurring~\cite{bowler_digitalscheduling_2022, guo_edgeuncertainty_2015, maceachren_visualsemiotics_2012}).
Although such techniques could be adapted to AR through methods like diminished reality~\cite{mori2017survey, chengdiminishreality}, they introduce inherent limitations: they reduce the visibility of the initially coarse-selected targets, and if the intended target is not recognized as the top candidate and blurred out, it impedes effective disambiguation. Furthermore, they are less generalizable to optical see-through devices. Thus, we focus on uncertainty visualization techniques that emphasize targets.

\section{Survey and Pointer \rev{Design}}
To explore and understand \UP{}s design possibilities, we conduct a systematic literature review to investigate annotation strategies and visual signifiers that could be used to add identity for disambiguation or to represent uncertain confidence levels in target selection scenarios.


Our systematic literature review follows PRISMA guidelines~\cite{moher2009prisma}. Our goals were to (1) identify prior research relevant to visualizing uncertainty during selection tasks, and (2) develop new possible visualization designs for uncertain target selection, extrapolated from these existing approaches. 

\subsection{Requirements and Sources}
Our review includes papers that meet the
following criteria:
\begin{enumerate}
    \item The paper must involve visualizations that communicate feedforward or uncertainty information directly to users.
    \item The paper needs to present a visualization technique for visible targets that shows area or volume. Specifically, we are looking for situated visualization~\cite{bressa_situatedvisualization_2022} with a 3D referent or embedded visualization~\cite{willett_embeddedviz_2017} with a 2D referent. 
\end{enumerate}
For Criterion 1, we excluded papers that address uncertainty solely from the perspective of robotic planning or model-internal computation, without a user-facing component. For example, we did not include techniques designed exclusively for ambiguity-aware robotic systems (e.g.,~\cite{schmidt_robotuncertainty_2008}) or predictive models (e.g.,~\cite{rogers_fingercloud_2010, zhang_atiimagelabeling_2024}).

Under Criterion 2, we required that the visualization techniques be applicable to target selection scenarios and capable of annotating real-world targets. This excluded abstract data visualizations such as density plots or quantile dotplots~\cite{kay_bus_2016,fernandes_quantiledotplots_2018}, which cannot be directly used for object-based visualization.

Across both criteria, we excluded works that communicate uncertainty without using visual modalities, such as approaches relying solely on auditory or haptic feedback ~\cite{jamshed_audionudges_2025, kunze_hapticseat_2018}.

\subsection{Survey Method}
\subsubsection{Phase 1: Identification}
We aimed to identify high-impact papers on visualization relevant to target selection scenarios. Given the interactive nature of our target scenario, we focused on venues that emphasize user interaction. We surveyed publications from the following proceedings: the ACM Conference on Human Factors in Computing Systems (CHI),
the ACM Symposium on Virtual Reality Software and Technology (VRST),
the ACM Symposium on User Interface Software and Technology (UIST),
the ACM Conference on Intelligent User Interfaces (ACM IUI),
the ACM Conference on Designing Interactive Systems (DIS),
the ACM Conference on Automotive User Interfaces and Interactive Vehicular Applications (AutomotiveUI),
the ACM Symposium on Spatial User Interaction (SUI), {the IEEE Transactions on Visualization and Computer Graphics (IEEE TVCG), the IEEE International Symposium on Mixed and Augmented Reality (IEEE ISMAR),} the IEEE Conference on Virtual Reality and 3D User Interfaces (IEEE VR), and 
the IEEE Symposium on 3D User Interfaces (IEEE 3DUI).

We focused our search on feedforward, uncertainty, and visualization, along with terms for our intended user
scenario (\ie{} in-situ, selection) via the advanced search fields of the (in-situ, selection) to appear in either the title or abstract.

\textsc{title: uncertainty OR feedforward OR (in-situ AND visuali*) OR (selection AND visuali*) OR}

\textsc{abstract: uncertainty OR feedforward OR (in-situ AND visuali*) OR (selection AND visuali*)}

We used structural boolean queries (not keyword search) via the advanced search field of IEEE and ACM Digital Libraries, covering full papers published between 1990 and 2025. The asterisk (*) was used as a wildcard to represent any number of unknown characters. This resulted in {721} results: {288 from TVCG}, 285 from CHI, {32 from ISMAR}, 30 from DIS, 27 from UIST, 24 from IEEE VR, 13 from AutomotiveUI, 8 from SUI, and 5 from VRST. We compiled the titles and abstracts of these 721 publications for screening in Phase 2.

\subsubsection{Phase 2: Screening}
We screened the titles and abstracts of the {721} papers collected in Phase 1 based on the inclusion criteria described above. 
{299} papers were selected for Phase 3, while {422} were excluded.
\subsubsection{Phase 3: Eligibility}
We reviewed the full-text articles for eligibility based on the two inclusion criteria. Papers were excluded at this stage if they did not meet either criterion upon closer full-text review or if they were not full papers. \rev{For example, we exclude non-archival articles, posters, workshop papers, and late-breaking work.} In total, {179} publications were excluded during this phase. 

\begin{figure} [ht]
    \centering
    \includegraphics[width=0.95\linewidth]{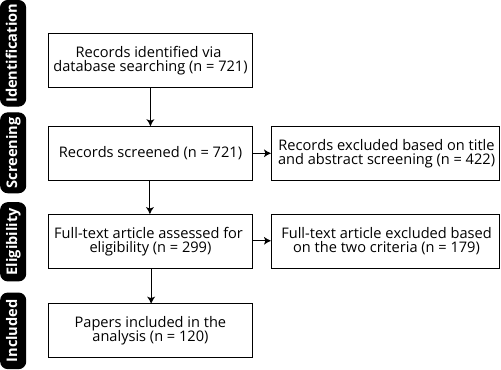}
    \caption{{Flow of information through the different phases of our systematic review, following PRISMA guidelines.}}
    \label{fig:PRISMA}
    \Description{Figure 2 presents a PRISMA-style flow diagram summarizing the systematic literature review process.
At the top, a box labeled “Records identified via database searching (n=721)” branches downward into subsequent screening stages.
The second box, “Records screened (n=721),” leads to a right-hand box indicating exclusion of 422 papers based on title and abstract.
The remaining 299 records move to full-text eligibility assessment, depicted in the next box. A right-hand branch shows 179 full-text articles excluded based on criteria.
The final bottom box indicates that 120 papers were included in the final dataset.
Arrows clearly mark the downward progression, and box alignment visually communicates decision flow, inclusion/exclusion criteria, and transparency of paper selection.}
\end{figure}

\begin{table*}[hbt]
    \caption{{Overview of systematic literature review, resulting in a dataset of 220 visualization techniques.}}
    \resizebox{0.9\textwidth}{!}{
\newlength{\Yw}
\setlength{\Yw}{\dimexpr(\textwidth-1.4cm-1.3cm-10\tabcolsep-0pt)/4\relax}
\newcolumntype{Y}{>{\centering\arraybackslash}p{\Yw}}

\newcolumntype{Z}{>{\centering\arraybackslash}p{1.3cm}}
\newcolumntype{G}{>{\centering\arraybackslash}p{1.4cm}}
\newcommand*{\mline}[1]{%
\begingroup
    \renewcommand*{\arraystretch}{1.1}%
    \begin{tabular}[c]{@{}>{\centering\arraybackslash}p{2cm}@{}}#1\end{tabular}%
\endgroup
}

\newcommand*{\mmline}[1]{%
\begingroup
    \renewcommand*{\arraystretch}{1.1}%
    \begin{tabular}[c]{@{}>{\arraybackslash}p{14cm}@{}}#1\end{tabular}%
\endgroup
}
{
\begin{tabular}{G *{4}{Y} |Z}
\toprule
\multicolumn{1}{l}{}  & \mline{\textit{\textbf{External}}}                                                                                                                                                                                                         & \mline{\textit{\textbf{Internal}}}                                                                                                                                                               & \mline{\textit{\textbf{Boundary}}}                                                                                     & \mline{\textit{\textbf{Fill}}}                                                                                                                                                                                                                                                                                                                                                                                                                                                                                                & \textit{\textbf{\%}}           \\
\end{tabular}
}
    }
    \label{tab:surey-result}
\end{table*}
\subsubsection{Coding Process}
We coded each of the {120} eligible papers along two dimensions: (1) \textbf{Pointer Archetype}, referring to the annotation and placement strategy used to associate the visualization with the referent (\eg{} adding a visible mark to the referent or outlining the referent’s boundary); (2) \textbf{Visual Signifier}, the visual property used to distinguish highlighted referents from each other or from non-referents (\eg{} color, size, etc.), which may or may not explicitly convey uncertainty.


Many papers include multiple visualizations within their use case and are therefore applicable to multiple categories. We apply multiple labels in such cases. 
For the eligibility and screening phase, two of the authors independently reviewed the entire collection. If either author marked a paper for inclusion, it was advanced to the next phase.
For dataset coding, the same two authors independently coded the final dataset and resolved any disagreements through discussion. The initial inter-rater agreement was {93\%}, with no discrepancies remaining after resolution.

\subsection{Survey Results}

\subsubsection{Pointer Archetype}
For the pointer archetype dimension, we discovered four primary categories that can describe most of our dataset: \f{external}, \f{internal}, \f{boundary}, and \f{fill}. Additionally, we include a fifth category, \f{others} for cases that do not fit into the main four and are considered unsuitable for AR target selection scenarios. \f{external} refers to visual annotations placed outside the referent’s area or volume, while \f{internal} refers to those placed and anchored within it. The \f{boundary} category includes annotations that follow or highlight the referent’s outline. \f{fill} applies a color or pattern change across the entire referent. 

Among the {120} papers of our dataset, we summarized a total of {220} visualization techniques, of which {16.36}\% belong to \f{external}, {26.37}\% \f{internal}, {18.18}\% \f{boundary}, and {39.09}\% are \f{fill}. Among the \f{others} category, some techniques use embodied virtual characters to express dialogue-based uncertainty around objects ~\cite{kim_bubbleu_2023, schmidt_naturalexpression_2024}, while others rely on blinking or animated motion cues ~\cite{ avdic_machinebody_2021, marquardt_airconstellations_2021, velloso_motioncorrelation_2021}. We consider these approaches overly distracting for on-the-go use, raising potential safety concerns and limiting their suitability for mobile or attention-sensitive AR scenarios.

\subsubsection{Visual Signifier}
We coded the dataset along the visual signifiers shown in previous literature~\cite{kunze_ardriveUV_2018, zeng_vizcollation_2023, kim_sensorviz_2022, kong_vizmapreview_2019, tamara_2015_vizAD}, which encapsulates 11 category, including {Position}, {Size}, {Length}, {Shape}, {Orientation}, {Color} (including saturation, hue, and luminance), {Texture}, {Opacity}, {Resolution}, {Text or Numeric}, and {Angle}.
{Among our dataset, {102} of the used signifiers are {color}, {35} are {size}, {26} are {opacity}, {17} are text, {12} are {texture}, {8} are {shape}, {6} are {position}, another {6} are {resolution}, {5} are {orientation}, and {3} are {length}.}

    \begin{figure*} [ht!]
        \centering
        \includegraphics[width=0.9\textwidth]{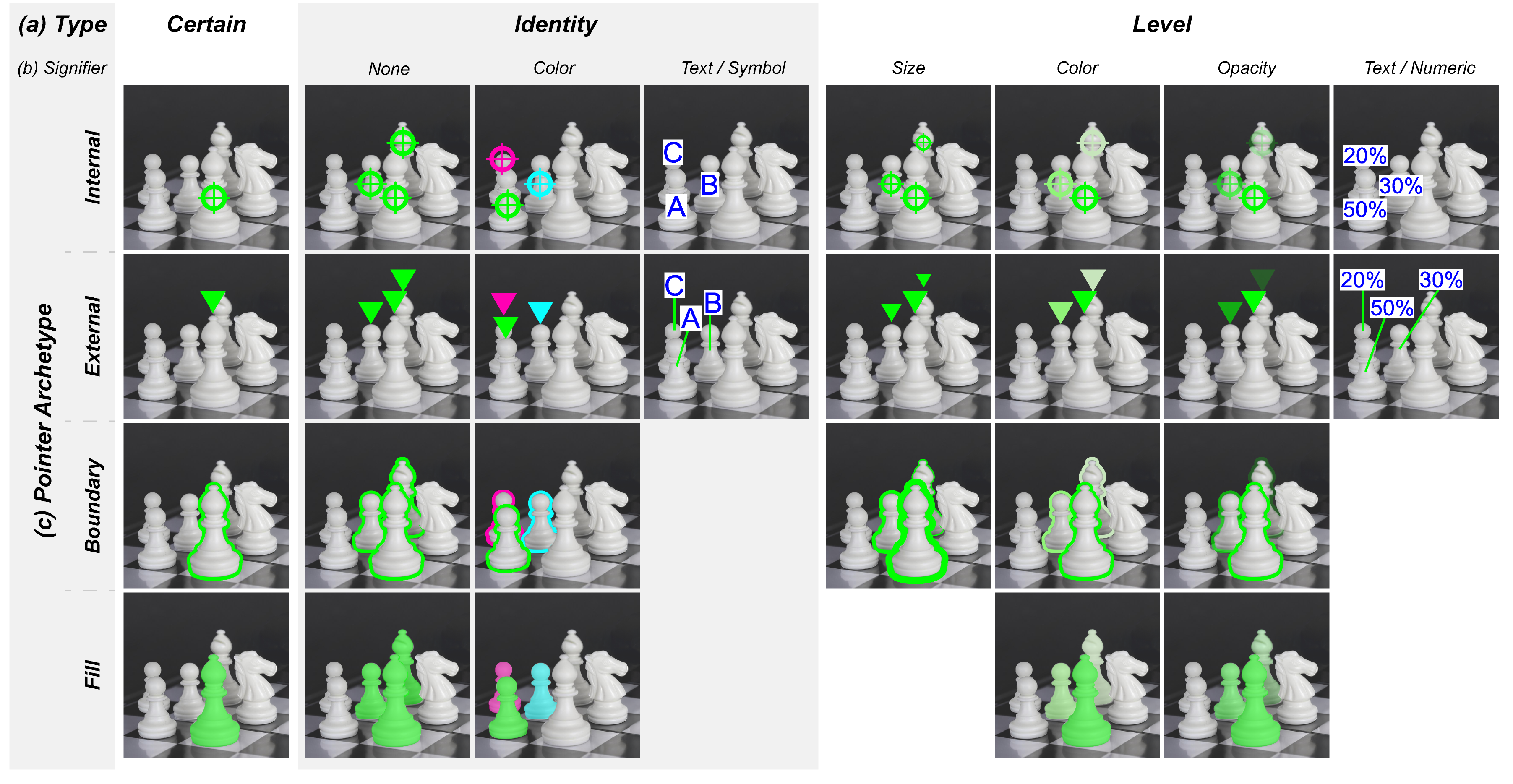}
        \caption{Pointer space of Uncertain Pointer, defined across three dimensions:
(a) {Uncertainty Complexity Type}, strategies for facilitating disambiguation that convey varying amounts of uncertainty information (\eg{} {Certain discloses no uncertainty, Identity shows the existence of uncertainty, Level reveals a graded level of uncertainty});
(b) {Visual} Signifier, visual attributes used to represent uncertainty levels or identities (\eg{} color, size, opacity, text/symbol);
(c) Pointer Archetype, different forms and spatial placements of visualizations relative to the target objects (\eg{} boundary, fill, internal, external).}
        \label{fig:design_space}
    \Description{Figure 3 is a multi-row grid showing the full design space of Uncertain Pointer across three dimensions: uncertainty complexity type, visual signifier, and pointer archetype.
Across the top, three uncertainty types (“Certain,” “Identity,” and “Level”) define the columns.
Along the left margin, the four pointer archetypes—Internal, External, Boundary, and Fill—define the rows.
Within each cell, example visualizations are displayed on grayscale scenes containing sets of 3D objects (e.g., chess pieces or figurines).
– Internal pointers place markers inside objects (e.g., small icons or localized overlays).
– External pointers use floating shapes or badges positioned near, but not overlapping, objects.
– Boundary pointers outline the silhouettes of each object with colored or glowing borders.
– Fill pointers apply color washes that tint the full surface of an object.
Identity signifier examples include distinct colors and text labels, while level pointers show gradations of opacity, size, or luminance that rank candidates.
Overall, the figure visually maps how combinations of pointer types and signifiers generate a space of 25 possible designs.}
    \end{figure*}
    
\subsection{\rev{Pointer Space}}





Based on our systematic literature review, we generated our pointer space based on the four pointer archetypes and the top four signifiers in our dataset (\eg{} {color}, {text}, {size}, {opacity});
also, with {uncertainty complexity}, which captures the level of uncertainty information conveyed by the visualization:
(1) \f{certain}: No uncertainty is expressed; only a single object is annotated, similar to a conventional deterministic pointer.
(2) \f{identity}: Visualizations that convey only the \textit{existence} of uncertainty and facilitate disambiguation; it annotates multiple candidate targets without encoding the magnitude of uncertainty between them. To support fast, easy disambiguation, {Identity} visualizations use visual identities (such as distinct {colors} or {text} labels). Since applying {size} or {opacity} changes inherently implies a ranked or leveled hierarchy, they are not included as identifiers in the {Identity} category. Also, we added a \f{none} signifier category with a uniform color to the {identity} pointer set, not only to set a baseline but also to showcase the presence of uncertainty.
(3) \f{level}: Visualizations convey graded uncertainty across multiple candidates using variations in visual signifiers (\eg{} intensity, size, or opacity), communicating the system's confidence in each candidate object being the intended selection target. In {level} pointer, for {Color} changes, we utilize a combination of increasing luminance and decreasing saturation~\cite{hengl2006maps}, based on ~\citet{maceachren_visualsemiotics_2012}'s and ~\citet{correll_errorbar_2014}'s results, as they both find it effective when (un)certainty is encoded using these properties.

{Overall, \UP{}'s pointer space is a set of \textit{visible}, \textit{overlaying}, \textit{abstract} AR visualization with purpose of \textit{directing attention} toward uncertain candidate set for target disambiguation (following~\citet{zollmann}'s AR visualization categorization), with three design dimensions: (1) uncertainty complexity type, (2) visual signifier, and (3) pointer archetype.}



\section{Online Experiment 1: Certain and Identity Visualizations}

The goal of this first online experiment is twofold: (1) to examine how different scenes, archetypes, and signifiers influence visualization effectiveness for Identity and Certain visualizations in terms of pointer identifiability while maintaining object visibility, and (2) to find low-performing or incompatible scene–archetype combinations to exclude from the follow-up study for level visualization. 

\subsection{Task and Procedure}
In the experiment, participants viewed a series of mock-up AR videos and completed object counting tasks and provided a collection of subjective ratings without performing target selection in the video scene. Each video featured an \UP visualization embedded within a specific real-world scene. After a tutorial explaining the simulated AR scenario and a brief practice session, participants proceeded to complete the tasks for each video trial. To assess participant attention levels, two attention checks were embedded intermittently throughout the study. Additionally, each participant was offered a 5-minute break after every quarter of the total number of trials to mitigate fatigue.
After completing all video trials, participants were asked to provide open-ended feedback and share their thoughts on the visualizations via text boxes.

We excluded data from any participants who failed either attention check, continuing recruitment until we reached 60 eligible participants. 12 participants failed our attention test in this study. This sample size was determined through our pre-registered power analysis to ensure 90\% statistical power for comparisons across \f{archetype}, \f{signifier}, and \f{archetype} \by{} \f{scene}.

\subsection{Preparation and Apparatus}
\begin{figure} [ht!]
        \centering
        \includegraphics[width=\linewidth]{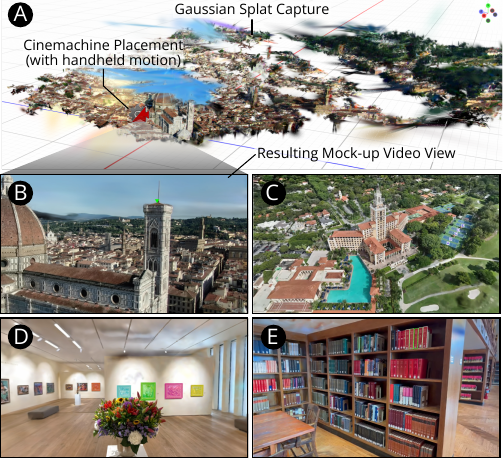}
        \caption{(A) Our mock-up AR video generation pipeline from Gaussian splat capture to resulting videos along with the 4 scenes used for our study with varying target distance and sparsity: (B) \f{sparse \by{} far}, shown with a \f{external}-\f{certain} pointer, (C) \f{dense \by{} far}, shown with \f{internal}-\f{text}-\f{identity} pointers, (D) \f{sparse \by{} near}, shown with \f{fill}-\f{color}-\f{identity} pointers, and (E) \f{dense \by{} near}, shown with \f{boundary}-\f{color}-\f{level} pointers, across the tower, tennis court, paintings, and books scenes, respectively. }
        \label{fig:scene-apparatus}
        \Description{Figure 4 contains a three-part pipeline diagram followed by four example scenes.
Panel (A) shows a Gaussian splat reconstruction of a real outdoor environment; the view resembles a point-cloud-like rendering with natural depth continuity.
A rightward arrow leads to a panel with a Unity Cinemachine camera icon, annotated with “with handheld motion,” indicating that naturalistic camera shake is added to simulate AR head movement.
The final panel shows a fully rendered AR mock-up video frame with pointer annotations embedded in the 3D scene.
Below the pipeline, four real-scene examples are shown:
– Sparse × Far: A distant courtyard with a few widely separated objects, annotated by a certain external pointer.
– Dense × Far: A cluster of tennis courts viewed from above, annotated with internal text-based identity pointers.
– Sparse × Near: A gallery-like scene with paintings annotated using color-based fill pointers.
– Dense × Near: A bookshelf scene where tightly packed books are annotated using level-based boundary pointers.
The figure demonstrates how the authors generated immersive, controlled AR stimuli for their user studies.}
    \end{figure}

The entire study protocol was implemented in {Qualtrics}, including the object-counting task and the subjective rating questions (Figure~\ref{fig:study-apparatus} A \& C).
The mock-up AR videos were generated in Unity. These videos cover four background scenes, combining two target distances (near \vs{} far) and two target sparsities (sparse \vs{} dense). Each background was created using a Gaussian splat capture (Figure~\ref{fig:scene-apparatus}), with 3D objects manually annotated for application of \UP's visualization effects. To simulate the sense of 3D motion from a first-person perspective, we used Unity’s Cinemachine to position the virtual camera and applied noise to the camera's motion to mimic a natural, continuous shake in an AR user’s perspective on the move. The resulting scenes were rendered as videos for use in the online study.

To annotate objects across pointer archetypes and visual signifiers, we followed a set of heuristics grounded in prior literature. The default pointer color was set to  
\raisebox{-0.1em}{\includegraphics[height=0.7em]{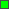}}~\f{green}, as it has been shown to offer high visibility in AR environments with varying backgrounds~\cite{merenda_colorandbg_2016, gabbard_colorblending_2-22} and is effective to convey a positive, confirmatory semantic meaning (or connotation) suitable for the most certain option~\cite{maceachren_visualsemiotics_2012}. For text-based signifiers, we adopted a ``billboard'' design~\cite{gabbard_textbillboard_2005}, using  \raisebox{-0.1em}{\includegraphics[height=0.7em]{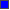}}~\f{deep blue} text 
on a \raisebox{-0.1em}{\includegraphics[height=0.7em]{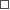}}~\f{white} background to ensure legibility across diverse real-world textures. To minimize visual clutter introduced by background interference, we also applied view management principles from~\citet{grasset_viewmanagementAR_2012}, avoiding edges and textured regions when placing external textual annotations. For identity pointers with the \f{color} signifier, we use easily verbalized colors with luminance similar to our default green to avoid visual hierarchy. {To accommodate users and participants with deuteranomaly or protanomaly (red-green color blindness), we applied colorblind-friendly adjustments by avoiding similar red-green channel intensities.
} The overall identity color set results in the use of \raisebox{-0.1em}{\includegraphics[height=0.7em]{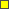}}~\f{yellow}, \raisebox{-0.1em}{\includegraphics[height=0.7em]{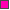}}~\f{pink}, \raisebox{-0.1em}{\includegraphics[height=0.7em]{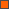}}~\f{orange}, and \raisebox{-0.1em}{\includegraphics[height=0.7em]{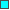}}~\f{blue}. 

\subsection{Study Design}
The study used a within-subject design and was pre-registered\footnote{\href{https://aspredicted.org/7h75-yxt5.pdf}{The link to Experiment 1 pre-registration}, revised during the review process to use repeated measures ANOVA with sphericity tests, along with post hoc t-tests using Holm–Bonferroni correction.} before deployment. It covered two of the \f{uncertainty complexities} in our pointer space: \f{certain} and \f{identity}, but these were not analyzed or compared to each other. Therefore, the independent variables were \f{archetype}, \f{signifier}, and \f{scene}:
\begin{itemize}
    \item 4 scenes with varying target distance and sparsity: \f{sparse \by{} far (sf)}, \f{dense \by{} far (df)}, \f{sparse \by{} near (sn)}, and \f{dense \by{} near (dn)}.
    \item 4 pointer archetypes: \EX, \IN, \BOUND, and \FILL.
    \item 3 signifiers for \f{identity}: \f{none}, \f{color}, and \f{text}.
\end{itemize}
Additionally, for each \f{archetype} \by{} \f{signifier} \by{} \f{scene} triad, we created 3 target configurations with varying numbers (3, 4, or 5) and target locations to minimize participants' learning effects across trials.
In total, we created 4 scenes \by 3 target variations \by (4 Certain Pointers + 10 Identity Pointers) = 168 video variations.
For each video trial, participants were asked to perform a counting task on how many targets in the video scene were being annotated with a visualization effect, referenced from the counting and identification task in previous visualization literature ~\cite{ hullman_inpursuitoferror_2019, sanyal_4uvcomparison_2009} and a series of subjective questions, including preference, target visibility using 1–7 continuous scales with thematic anchors. {For workload-related measures, we only used a subset of the NASA-TLX survey that was relevant for our task, while mitigating fatigue across the repeated trials.} 
The dependent variables are shown as follows:
\begin{itemize}
    \item \m{Duration (s)}: From video onset to counting task submission.
    \item \m{Error in Count}: Absolute difference from the correct number of targets.
    \item \m{Confidence}: “How confident are you about your answers?” (1 = Very Unsure, 7 = Very Confident)
    \item \m{Mental Ease}: “How much mental effort was required to perform the tasks?” (1 = Very High, 7 = Very Low)
    \item \m{Target Visibility}: “To what extent do you think the visualization blocks the user's view, making it difficult to see the details of the objects?” (1 = Very Much, 7 = Very Little)
    \item \m{Preference}: “How likely are you to choose to use this visualization during this scenario?” (1 = Very Unlikely, 7 = Very Likely)
\end{itemize}
{We adopted continuous rating scales instead of ordinal Likert scales. This allows us to perform more flexible and rigorous analyses with interval-level statistics (\eg{} mean comparison). }

Each participant was presented one of the three target variation for each video trial, so a total of 56 unique video trials, covering all combinations of annotation \f{archetype} \by{} \f{signifier} \by{} \f{scene} for \f{identity} visualization and \f{archetype} \by{} \f{scene} for \f{certain} visualization. The trials were organized into four blocks, each containing 14 shuffled video trials with the same scene. The order of the scenes across blocks was also randomized.

\subsection{Participants}
We recruited 60 participants (ages 18 to 52, \mean{23.1}, \sd{5.5}) via mailing lists and word of mouth. 36 identified as women, 20 as men, 3 as non-binary, and 1 preferred not to disclose. 
Only 1 participant reported being colorblind. 
All participants provided informed consent and were compensated with a \$15 Amazon gift card. The study took less than 1 hour to complete.

\subsection{Data Analysis}
We analyzed behavioral and subjective measures using repeated measures ANOVA and, when sphericity was violated, reported Greenhouse-Geisser-corrected degrees of freedom, F-statistics, and p-values. For post hoc comparisons, we conducted paired t-tests with the Holm–Bonferroni correction. The analyses were conducted across the independent variables (\f{scene}, \f{archetype}, and \f{signifier}). Specifically, we conducted comparisons among \f{archetypes} for both \f{certain} and \f{identity} pointers, and among \f{signifiers} and \f{archetype} \by{} \f{scene} for \f{identity} pointers. 

Given the large number of visualization conditions and the many possible groupings for pairwise comparisons, we focus on reporting results for \f{archetype} and \f{signifier} comparisons, while only presenting a few specific findings for \f{archetype} \by{} \f{scene} (Figure~\ref{fig:uniform_barplot}); full statistical result are reported in the Appendix. We complement our quantitative results with participants’ qualitative feedback from open-ended responses to provide possible explanations.

Additionally, we performed K-means clustering on combinations of \f{archetype} \by{} \f{signifier} \by{} \f{scene} in the \f{identity} pointer condition to rule out potentially less effective pairings before proceeding to Study 2 with \f{level} pointers.

\subsection{Results for Certain Pointers}
\begin{figure*} [ht!]
        \centering
        \includegraphics[width=0.85\textwidth]{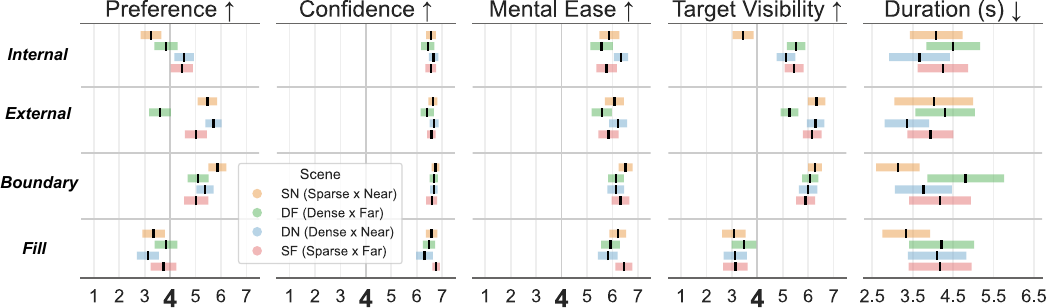}
        \caption{Study results for \f{certain} pointers in terms of \m{Preference}, \m{Confidence}, \m{Mental Ease}, \m{Target Visibility}, and \m{Duration} for each \f{archetype}, \f{scene}, and their combinations. The error bars represent 95\% confidence intervals. All participants answered correctly on the counting task, so \m{Error in Count} is always 0 and excluded from the plot.}
        \label{fig:certain_barplot}
        \Description{Figure 5 contains five horizontally aligned scatterplot panels showing results for Certain pointers across four archetypes: Internal, External, Boundary, and Fill.
Each subplot visualizes one dependent measure—Preference, Confidence, Mental Ease, Target Visibility, and Duration—across four scene layouts (Sparse × Far, Dense × Far, Sparse × Near, Dense × Near).
Data points are jittered horizontally, color-coded by scene condition, and include 95
The Boundary pointer consistently appears at the rightmost (higher) ends of Preference, Mental Ease, and Target Visibility scales. Internal and Fill pointers cluster toward lower ratings, especially in visibility measures where occlusion is apparent.
Duration values are relatively stable across archetypes, with small increases in near/dense scenes.
The figure visually communicates the strong advantage of boundary-based highlighting for clarity and non-occlusion during single-target annotation.}
    \end{figure*}
{We first discuss the result for \f{certain} pointer, where there's only one object being annotated.} 
For \f{archetype} comparison in \f{Certain} pointers, we observed significant effects in  \m{Preference} ({\rmanova{3}{177}{50.48}{.0001}{0.46}}), \m{Confidence} ({\rmanova{3}{177}{3.34}{.05}{0.053}}), \m{Mental Ease} ({\rmanova{3}{177}{7.27}{.0005}{0.11}}), and \m{Target Visibility} ({\rmanova{1.68}{99.17}{122.12}{.0001}{0.67}, Greenhouse-Geisser-corrected}) with {repeated measures} ANOVA.

As shown in Figure~\ref{fig:certain_barplot} and revealed by {post hoc pairwise t-tests with Holm–Bonferroni correction}, across all scenes, the \BOUND{} archetype consistently yields the best ratings. It received significantly higher ratings than \IN{} ({\p{.0001}, Cohen's \dvalue{1.71}}), \f{external} ({\p{.005}, \dvalue{0.41}}), and \FILL{} ({\p{.0001}, \dvalue{1.14}}) in \m{Preference}; surpassed \f{internal} ({\p{.005}, \dvalue{0.49}}), \EX{} ({\p{.005},\dvalue{0.5}}), and \f{fill} ({\p{.1}, \dvalue{0.31}}) in \m{Mental Ease}; and surpassed \f{internal} ({\p{.0001}, \dvalue{1.39}}) and \f{fill} ({\p{.0001}, \dvalue{1.65}}) in \m{Target Visibility}. 35\% of the participants mentioned they favor \f{boundary} as their most preferred archetype, as it does not \pquote{obscure}{p7}, \pquote{block}{p10}, or \pquote{obstruct}{p26} the items and \pquote{kept the object in clear view}{p41}. Participants also express that it's easy to \pquote{identify}{p42} and \pquote{spot}{p8}.

\f{external} is considered the second-best option except for the \f{df} (dense and far) scene. It received significantly higher overall ratings than \f{internal} ({\p{.0001}, \dvalue{0.93}}) and \f{fill} ({\p{.0001}, \dvalue{0.96}}) in \m{Preference}; and surpassed \f{internal} ({\p{.0001}, \dvalue{1.14}}) and \f{fill} ({\p{.0001}, \dvalue{1.57}}) in \m{Target Visibility}. 25\% of participants expressed a preference for the \f{external} archetype, quoting its ability to \pquote{allow me to see the objects}{p55} and being the \pquote{least occlusive}{p46}. However, participants also noted limitations of \f{external} in the \f{df} scene. Specifically, when pointing to small, cluttered tennis courts, external pointers were described as \pquote{confusing}{p56} and prone to \pquote{being mistaken}{p37} as the pointer is displaced from the objects.

\f{internal} and \f{fill} are the less preferred archetypes. Among the two, \f{internal} significantly outperformed \f{fill} in \m{Preference} ({\p{.001}, \dvalue{0.35}}) and \m{Target Visibility} ({\p{.0001}, \dvalue{1.17}}). However, participants also mentioned they are more useful for \pquote{faraway objects}{p54}. 

\subsection{Results for Identity Pointers}
    \begin{figure*} [ht!]
        \centering
        \includegraphics[width=\textwidth]{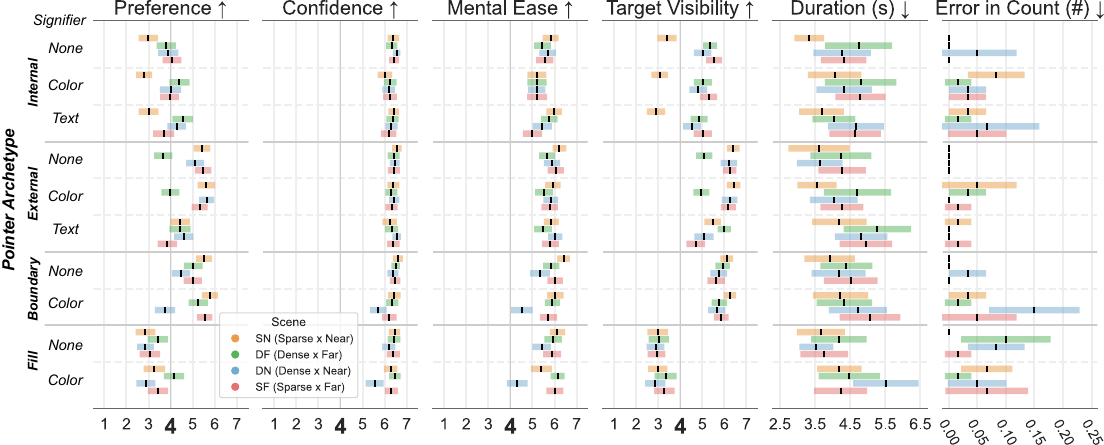}
        \caption{Study results for \f{identity} pointers in terms of \m{Preference}, \m{Confidence}, \m{Mental Ease}, \m{Target Visibility}, \m{Duration}, and \m{Error in Count} for each \f{archetype}, \f{signifier} \f{scene}, and their combinations. The error bars represent 95\% confidence intervals.\looseness=-1}
        \label{fig:uniform_barplot}
        \Description{Figure 6 is a large grid of labeled scatterplots showing Identity-pointer results. Rows correspond to pointer archetypes (Internal, External, Boundary, Fill). Columns represent signifiers (None, Color, Text).
Within each small plot, data points show subjective ratings (Preference, Confidence, Mental Ease, Target Visibility) and objective metrics (Duration, Error in Count).
Boundary pointers generally dominate across Preference, Mental Ease, and Visibility, except in Dense × Near, where multi-object occlusion reduces performance.
External pointers perform second-best overall, while Fill pointers perform worst due to heavy occlusion. The “None” signifier yields lower preference but higher counting accuracy.
The grid highlights how signifier choice interacts with pointer archetype and scene complexity.}
    \end{figure*} 
{Here, we discuss the result for \f{identity} pointer, where multiple targets in the uncertainty set are annotated with either the same color (\ie{} \f{none}), distinct hues of color (\ie{} \f{color}), or a text label (\ie{} \f{text}), without level graduation.}
\subsubsection*{\textbf{Archetype}}
For \f{archetype} comparison in \f{identity} pointers, we observed significant effects in \m{Preference} ({\rmanova{3}{177}{74.38}{.0001}{0.56}}), {\m{Confidence} (\rmanova{2.24}{132.25}{2.95}{.0001}{0.05})}, \m{Mental Ease} ({\rmanova{3}{177}{9.77}{.0001}{0.14}}), \m{Target Visibility} ({\rmanova{1.57}{92.51}{147.07}{.0001}{0.71}}), and \m{Error in Count} ({\rmanova{2.09}{123.38}{3.62}{0.05}{0.06}}) with {repeated measures} ANOVA.

As shown in Figure~\ref{fig:uniform_barplot} and revealed by {post hoc pairwise t-tests with Holm–Bonferroni correction}, the general trend of \f{identity} pointers follows its \f{certain} counterparts. For example, \BOUND{} is still the top-performing archetype, similar to its single-pointer version. It received significantly higher ratings than \IN{} ({\p{.0001}, Cohen's \dvalue{1.23}}), \EX{} ({\p{.05}, \dvalue{0.32}}), and \FILL{} ({\p{.0001}, \dvalue{1.40}}) in \m{Preference}; surpassed \f{internal} ({\p{.005}, \dvalue{0.48}}) in \m{Mental Ease}; exceed \f{internal} ({\p{.0001}, \dvalue{1.22}}) and \f{fill} (\p{.0001}, \dvalue{1.74}) in \m{Target Visibility}; {had lower error than \f{external} (\p{0.1}, \dvalue{0.3}) in \m{Error in Count}}. 
However, it received significantly lower ratings in \f{dn} scenes, which feature densely placed objects at near distances (Appendix~\ref{sec:appendix-experiment1}). This suggests that multi-object annotation has different design considerations from a single one.

\f{external} is still considered the second-best preferred option, similarly, except for the \f{df} (Dense and Far) scene. It received significantly higher ratings than \f{internal} ({\p{.0001}, Cohen's \dvalue{1.17}}) and \f{fill} ({\p{.0001}, \dvalue{1.25}}) in \m{Preference}; {also, had higher ratings than \f{internal} (\p{.1}, \dvalue{0.35}) and \f{fill} (\p{.1}, \dvalue{0.32}) in \m{Confidence}}; received higher ratings compared to \f{internal} ({\p{.0001}, \dvalue{0.62}}) and \f{fill} ({\p{.05}, \dvalue{1.33}}) in \m{Mental Ease}; exceeded \f{internal} ({\p{.0001}, \dvalue{1.34}}) and \f{fill} ({\p{.0001}, \dvalue{1.76}}) in \m{Target Visibility}; {lastly, it have higher ratings (\ie{} lower error) versus \f{internal} (\p{.05}, \dvalue{0.34}), {\f{boundary} (\p{.1}, \dvalue{0.3})}, and \f{fill} (\p{.01}, \dvalue{0.43}) in \m{Error in Count}}.

\subsubsection*{\textbf{Signifier}} 
{Repeated measures} ANOVA revealed significant effects of  visual signifier on \m{Preference} ({\rmanova{2}{118}{4.27}{.05}{0.07}}), {\m{Confidence} (\rmanova{2}{118}{3.30}{.05}{0.05}), \m{Mental Ease} (\rmanova{2}{118}{11.07}{.0001}{0.16}), \m{Duration} (\rmanova{2}{118}{8.17}{.0005}{0.12})}, and \m{Error in Count} ({\anova{2}{118}{3.34}{.05}{0.05}}).

For \m{Preference}, participants significantly favored \f{color} over \f{text} ({\p{.05}, \dvalue{0.26}}). Several participants \textit{[p1, p8, p10, p12, p22, p57]} noted that colors made it easier to distinguish targets and were more immediately recognizable than text labels.

However, for \m{Error in Count}, \f{none} had lower error than \f{color} ({\p{.1}, \dvalue{0.29}}), with participants noting that \pquote{The same color is easier to count.}{p43} due to visual consistency.
\f{text} also surpassed \f{color} ({\p{.0001}, \dvalue{0.72}}) for counting accuracy. {This superiority also reflects in \m{Mental Ease}, where \f{none} had higher ratings than \f{color} (\p{.0001}, \dvalue{0.61}) and \f{text} (\p{.05}, \dvalue{0.28}), and in \m{Duration}, where \f{none} required less time than \f{color} (\p{.005}, \dvalue{0.46}) and \f{text} (\p{.005}, \dvalue{0.46}). Additionally, \f{none} showed higher ratings in \m{Confidence} than \f{color} (\p{.05}, \dvalue{0.34}).}

These findings highlight a tradeoff: while \f{color} is preferred for visual clarity and verbal selection, it may impair quick counting in pointer recognition \vs{} more uniform signifiers (\ie{} \f{none}).

\subsection{Recommendations Based on Clustering}
    \begin{figure} [ht!]
        \centering
        \includegraphics[ width=\columnwidth]{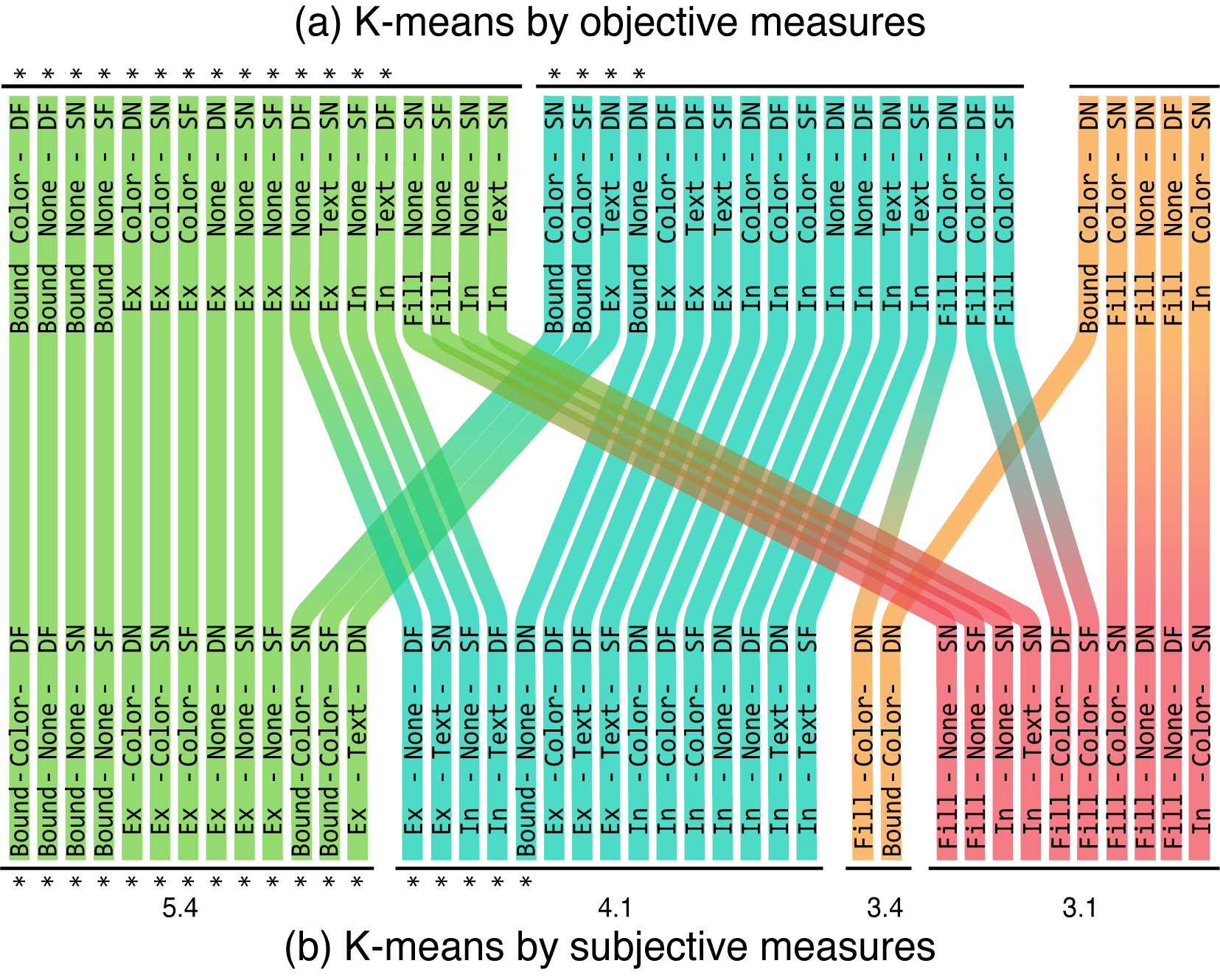}
        \caption{\f{archetype}-\f{signifier}-\f{scene} Groups.
(a) Clusters based on object-level metrics. (b) Clusters organized by subjective measures, with mean user preference displayed for each group. The asterisk (*) indicates the recommended best-performing group, demonstrating strong \f{archetype}-\f{signifier}-\f{scene} synergy. These specific \f{archetype}-\f{scene} pairs were selected to advance to Study 2, where we evaluate \f{level} pointer visualizations.}
        \label{fig:uniform_clustering}
        \Description{Figure 7 contains two cluster visualizations summarizing high-level performance patterns.
Panel (a) clusters Archetype × Signifier × Scene combinations based on subjective measures (Preference, Confidence, Mental Ease, Target Visibility). Each cluster is represented by a color-coded bar whose height indicates mean preference. Low-performing clusters (dominated by Fill archetype) appear grouped separately from higher-performing groups (primarily External and Boundary).
Panel (b) clusters the same combinations using objective metrics (Duration, Error in Count). Bars illustrate centroid-level performance differences.
Asterisks indicate clusters selected for inclusion in Study 2 (Level pointers), reflecting high synergy between visual clarity and performance.
The figure communicates how multidimensional metrics were used to filter design candidates systematically.}
    \end{figure} 

To summarize our findings across multiple metrics and reconcile potential conflicts between subjective and objective evaluations{, similar to previous gesture research ~\cite{cami_unimanual_2018, tsai_gg_2024}}, we applied K-means clustering to each \f{archetype} \by{} \f{scene} \by{} \f{signifier} triad based on metric scores. Given the observed divergence between subjective metrics (such as \m{Target Visibility} vs objective error-related measures), we generated two separate sets of clusters using distinct feature vectors. 
{Across 50 random initializations, clustering results were stable. 
All pairwise Adjusted Rand Index (ARI) values exceeded 0.95. The final model (yielding the lowest WCSS) resulted in 3 and 4 clusters for objective and subjective measures, respectively, with consistent centroid positions across runs. The overall silhouette coefficient was 0.66 and 0.52, indicating well-separated clusters.}

The first clustering set contains objective performance measures, including \m{Duration} and \m{Error in Count}. The second clustering set focuses on subjective metrics: \m{Preference}, \m{Confidence}, \m{Mental Ease}, and \m{Target Visibility}. For subjective clusters, we used the mean cluster-level \m{Preference} score to establish a relative group ordering. For objective clusters, ordering was based on statistical test results and the mean differences.

For subjective metrics, there was cluster separation between \FILL{} archetype \vs{} the others, suggesting the former are perceived to be less preferable. The \f{fill} and \IN{} archetypes with \f{sn} and \f{sf} scenes showed the largest shift between ordered groups, but all other pairings shifted no more than one adjacent group.

Based on the clustering results and main findings from Study 1, we excluded the \f{fill} archetype from our follow-up experiment on \f{level} visualizations. Specifically, we selected \f{archetype}-\f{scene} combinations that appeared in the top-performing clusters of either subjective or objective measures, and were not part of the lowest-performing cluster in either group. Among the combinations that fell into the second-tier cluster for both metrics, we included only the \f{boundary}-\f{dn} pair, as it outperformed top conditions in multiple measures, which are \f{external}-\f{dn} in \m{Error in Count} (\p{.05}) and \f{internal}-\f{far} in \m{Target Visibility} (\p{.1}). Overall, we include all \f{scene} conditions with \EX{} and \BOUND{} pointers and the two scenes with further targets (\f{SF} and \f{df}) with \f{internal} pointers for Experiment 2.

\section{Online Experiment 2: Level Visualizations}
The goal of our second online experiment is to examine how different scenes, archetypes, and signifiers influence visualization effectiveness for \f{level} visualizations.

\subsection{Preparation and Apparatus}
This study utilizes the same four Gaussian splat captures for rendering AR mock-up videos, and 
Qualtrics for study protocol.

For each change in visual signifier, we based our heuristic on Steven's power law~\cite{stevens_powerlaw_1970} (${\displaystyle \psi (I)=kI^{a}}$) to determine the intensity and changes for \f{color}, \f{opacity}, and \f{size} signifiers. The exponents $a$ we used are 1.8, 1.7, and 1, respectively, while $k$ is fitted to maintain minimum visibility for each pointer within each scene, based on previous visual research in transparency~\cite{bartram_transparencymin_2011}, saturation (for \f{color}) ~\cite{saturation_threshold}, and size~\cite{oh_size_2020}. 

For each target variation, we assigned the most visible pointer in the scene to the maximum value and the least visible to the minimum (as defined by the above-mentioned visibility consideration). The intermediate pointer intensities ($I$) were then linearly interpolated. As a result, variations with fewer targets exhibited more pronounced visual differences between candidates. 
To account for practical opacity levels of AR and smart glasses, {we conducted a small survey of the commercial optical see-through AR headsets and glasses with recent tint and dimming technology regarding their opacity level\footnote{Opacity Level Survey: \href{https://developer-docs.magicleap.cloud/docs/guides/features/dimmer-feature/}{Magic Leap 2} (80-100\%), \href{https://us.shop.xreal.com/products/xreal-one-pro}{Xreal One/One Pro} (100\%, near black out), \href{https://www.rayneo.com/products/rayneo-air-3s-xr-glasses}{RayNeo Air 3S} (99.6\%),  \href{https://www.rayneo.com/products/rayneo-air-3s-xr-glasses}{Snap Spectacles 5} using AlphaMicron’s E-Tint (84\%).}, which revealed a range of 80-100\%. Combined with common legal safety standards for light-blocking glasses in most countries and states (Category 3, 8\%-18\%)\footnote{Sunglass standards and category information: \href{https://cdn.standards.iteh.ai/samples/77321/f2bad2e34f7a4f4aaee205c3fc560ebf/ISO-12312-1-2022.pdf}{ISO 12312-1} and \href{https://webstore.ansi.org/standards/vc\%20(asc\%20z80)/ansiz802018?source=blog}{ANSI Z80.3}.}, accordingly, we set the upper bound of all visualizations' opacity to 90\%.

\subsection{Design, Task, and Procedure}
This study followed a similar format to our first online study but differed in the visualization conditions and the task design. Specifically, it introduced an additional identification task referenced from the searching task in previous uncertainty visualization literature ~\cite{vsup, hullman_inpursuitoferror_2019, sanyal_4uvcomparison_2009}  where participants selected the targets labeled by the pointers with most and least certainty levels {in the mock-up video scene using their mouse} (Appendix A, Figure~\ref{fig:study-apparatus} B). They also answered an added subjective question evaluating the \m{Intuitiveness and Logic} of the uncertainty signifiers, which is commonly tested for uncertainty visualization ~\cite{boukhelifa_sketchiness_2012, maceachren_visualsemiotics_2012}.

The study was also a within-subject design, which was pre-registered\footnote{\href{https://aspredicted.org/gss2-t59j.pdf}{The link to Experiment 2 pre-registration}, revised during the review process to use repeated measures ANOVA with sphericity tests, along with post hoc t-tests using Holm–Bonferroni correction} before deployment. The target number of users, based on our power analysis to achieve 90\% power for analyzing \f{archetype}, \f{signifier}, and \f{archetype} \by{} \f{scene}, was 40. The independent variables were \f{archetype}, \f{signifier}, and \f{scene}:
\begin{itemize}
    \item 4 Scenes with \f{sparse} \by{} \f{far (sf)}, \f{dense} \by{} \f{far (df)}, \f{sparse} \by{} \f{near (sn)}, and \f{dense} \by{} \f{near (dn)} (Near scenes are only for \f{external} and \f{boundary}).
    \item 3 Pointer Archetypes: \EX{}, \IN{}, \BOUND{}
    \item 4 Signifiers: \f{color}, \f{opacity}, \f{size}, and \f{text} (Text is only for External and Internal)
\end{itemize}
Similar to Study 1, for each \f{archetype} \by{} \f{signifier} \by{} \f{scene} triad, we created 3 target configurations with varying numbers (3, 4, or 5) and locations of targets to minimize participants' learning effect across trials.
In total, we created 108 video variations. Each participant was presented with one of the three target variations for each video trial, for a total of 36 unique video trials, covering all selected combinations of annotations \f{archetype} \by{} \f{signifier} \by{} \f{scene} for \f{level} visualizations.

Each video trial yields the same dependent variables, including \m{Duration (S)}, \m{Error in Count}, \m{Mental Ease}, \m{Target Visibility}, and \m{Preference}, with the duration spanning from the onset of the video to submitting the task. This study yields additional dependent variables, including: 
\begin{itemize}
    \item \m{Error in Most}: Absolute number of level difference from the correct most certain target to participant's answer.
    \item \m{Error in Least}: Absolute number of level difference from the correct least certain target.
    \item \m{Intuitiveness / Logic}: ``How intuitive or easy was it to map the visualization design to the system's confidence level across the candidate set?'' (1 = Very Difficult, 7 = Very Easy)
\end{itemize}

\subsection{Participants}
We recruited 40 participants (ages 18 to 46, \mean{21.9}, \sd{5.4}) via mailing lists and word of mouth, with no returning participants from the first experiment. 28 identified as women and 12 as men.
None of the participants reported being colorblind.
All participants provided informed consent and were compensated with a \$15 Amazon gift card. The study took less than one hour to complete.

\subsection{Results for Level Pointers}

\begin{figure*} [ht!]
    \centering
    \includegraphics[width=0.9\textwidth]{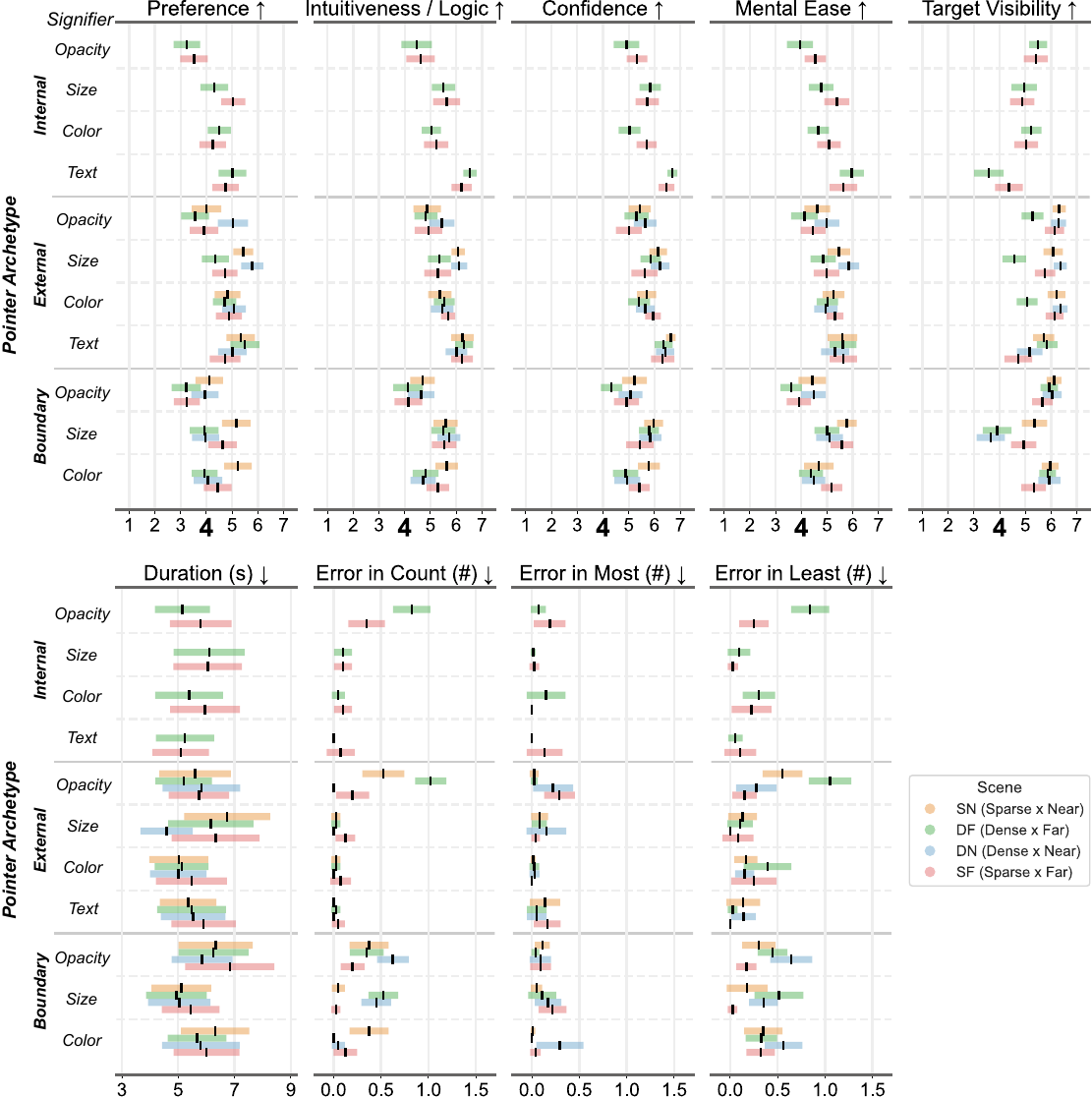}
    \caption{Study results for \f{level} pointers in terms of \m{Preference}, \m{Intuitiveness / Logic}, \m{Confidence}, \m{Mental Ease}, \m{Target Visibility}, \m{Error in Count}, \m{Error in Most}, \m{Error in Least}, and \m{Duration}, for each \f{archetype}, \f{signifier} \f{scene}, and their combinations. The error bars represent 95\% confidence intervals.}
    \label{fig:level_barplot}
    \Description{Figure 8 comprises a densely packed grid of scatterplots showing Level-pointer outcomes across archetypes (Internal, External, Boundary) and signifiers (Color, Opacity, Size, Text).
Top rows show subjective metrics: Preference, Intuitiveness/Logic, Confidence, Mental Ease, and Target Visibility.
Bottom rows show performance metrics: Duration, Error in Count, Error in Most-Likely Target, Error in Least-Likely Target.
External pointers generally appear with the highest Preference and Intuitiveness scores, with tight clustering indicating consistent interpretation across scenes. Boundary pointers show low intuitiveness and higher error rates, especially for least-likely target identification.
Signifier patterns vary:
– Color and Size show strong intuitive gradients.
– Opacity reduces clutter but sometimes makes levels harder to distinguish.
– Text is precise but slower due to reading demands.
The figure highlights nuanced trade-offs between interpretability, cognitive load, and error performance across spatial layouts.}
\end{figure*} 
{We discuss the result for \f{level} pointers, where multiple targets in the uncertainty set are annotated with either \f{color}, \f{size}, \f{text}, or \f{opacity} with variable intensities.}
\subsubsection*{\textbf{Archetype}}
For \f{archetype} comparison, significant effects were found across multiple measures using {repeated measures ANOVA}, including \m{Preference} ({\anova{2}{78}{10.91}{.0001}{0.22}}), \m{Intuitiveness / Logic} ({\rmanova{2}{78}{14.05}{.0001}{0.27}}), \m{Confidence} ({\rmanova{2}{78}{12.96}{.0001}{0.3}}), \m{Mental Ease} ({\rmanova{2}{78}{7.08}{.005}{0.15}}), \m{Target Visibility} ({\rmanova{2}{78}{22.89}{.0001}{0.37}}), {\m{Error in Count} (\rmanova{2}{78}{10.38}{.0001}{0.21})}, and \m{Error in Least} ({\rmanova{2}{78}{9.12}{.0005}{0.19}}), but not \m{Error in Most}.

{Overall, \EX{} received the highest ratings across multiple measures, while \BOUND{} received the lowest. In \m{Preference}, \f{external} surpassed both \f{boundary} (\p{.0001}, \dvalue{0.73}) and \IN{} (\p{.001}, \dvalue{0.60}). In \m{Intuitiveness / Logic}, \f{external} surpassed both \f{boundary} (\p{.0001}, \dvalue{0.94}) and \f{internal} (\p{.05}, \dvalue{0.37}), while \f{boundary} received the lowest scores, lower than \f{internal} (\p{.05}, \dvalue{0.44}). A similar trend is also observed in \m{Confidence}, where \f{external} has higher ratings \vs{} \f{boundary} (\p{.0001}, \dvalue{0.90}) and \f{internal} (\p{.1}, \dvalue{0.27}), while \f{internal} surpassed \f{boundary} (\p{.005}, \dvalue{0.58}). For \m{Mental Ease}, \f{boundary} received lower ratings compared to \f{external} (\p{.001}, \dvalue{0.63}) and \f{internal} (\p{.1}, \dvalue{0.34}).}

Error-related measures further confirmed this observation. {For \m{Error in Count}, \f{external} (\p{.0001}, \dvalue{0.84}) and \f{internal} (\p{.1}, \dvalue{0.29}) were both lower than \f{boundary}, while \f{external} yielded even lower error than \f{internal} (\p{.05}, \dvalue{0.4}). Similarly, in \m{Error in Least}, \f{boundary} pointers yielded higher error than both \f{external} (\p{.0005}, \dvalue{0.67}) and \f{internal} pointers (\p{.01}, \dvalue{0.50}).}

{The only measures where \f{boundary} was not rated as the worst is in \m{Target Visibility}, where \f{internal} pointers were rated lowest, lower than both \f{boundary} (\p{.005}, \dvalue{0.54}) and \f{external} (\p{.0001}, \dvalue{0.99}). Nevertheless, \f{boundary} was still lower than \f{external} in visibility (\p{.0005}, \dvalue{0.63}).}

Overall, \f{external} pointers was the best. Then, \f{internal} was advantaged over \f{boundary} in most measures except for target visibility, whereas participants rated \f{boundary} as the worst archetype across most measures.  For example, \pquote{I strongly disliked the `border' visualizations.}{p39}
Compared to our results with \f{certain} and \f{identity} pointers, \f{boundary} shows poor synergy with graded visual signifiers. We believe this is due to its thinner visible area, which limits the perceptibility of intensity-based visual changes.

\subsubsection*{\textbf{Signifier}}
For \f{signifier} comparison, significant effects emerged across all subjective measures and also objective \m{Error in Count} and \m{Error in Least}.

Overall \m{Preference} showed significant effects ({\rmanova{1.95}{76.04}{16.22}{.0001}{0.29}}). \f{opacity} was rated lower than \f{color} ({\p{.0001}, \dvalue{0.93}}), \f{text} ({\p{.0001}, \dvalue{0.8}}), and \f{size} ({\p{.0001}, \dvalue{1.01}}).

In \m{Intuitiveness / Logic} ({\rmanova{3}{117}{37.53}{.0001}{0.50}}), \f{opacity} again performed worst, receiving significantly lower ratings compared to \f{size} ({\p{.0001}, \dvalue{1.01}}), \f{color} ({\p{.0005}, \dvalue{0.72}}), and \f{text} ({\p{.0001}, \dvalue{0.99}}). \f{color} also rated below both \f{size} ({\p{.005}, \dvalue{0.53}}) and \f{text} ({\p{.0001}, \dvalue{0.99}}). {\f{text} was rated the highest, even compared to \f{size} (\p{.0005}, \dvalue{0.66}).}

Similarly, in \m{Confidence}, {repeated measures} ANOVA revealed a significant main effect ({\rmanova{3}{117}{47.13}{.0001}{0.55}}). Post hoc comparisons showed that \f{opacity} was rated significantly lower than \f{size} ({\p{.0001}, \dvalue{1.18}}), \f{color} ({\p{.005}, \dvalue{0.49}}),  and \f{text} ({\p{.0001}, \dvalue{1.37}}), while \f{color} was also rated lower than \f{size} ({\p{.0005}, \dvalue{0.65}}) and \f{text} ({\p{.0001}, \dvalue{1.27}}). {\f{text}, again, was rated the highest, even compared to \f{size} (\p{.0001}, \dvalue{0.79}).}

The same trend was also observed in \m{Mental Ease} ({\rmanova{1.83}{   71.29}{23.94}{.0001}{0.38}}), \f{text} surpassed \f{opacity} ({\p{.0001}, \dvalue{1.02}}), {\f{size} (\p{.1}, \dvalue{0.28}), and \f{color} (\p{.005}, \dvalue{0.55})}, while \f{size} surpassed \f{opacity} ({\p{.0001}, \dvalue{1.31}}) and \f{color} ({\p{.005}, \dvalue{0.60}}). {\f{opacity} was rated the lowest, even compared to \f{color} (\p{.0001}, \dvalue{0.85}).}

{Error-related behavioral measures further confirmed the same trend as significant main effects were found in \m{Error in count} (\rmanova{2.07}{80.85}{79.10}{.0001}{0.67}) and \m{Error in Least} (\rmanova{3}{117}{45.55}{.0001}{0.55}), revealed by repeated measures ANOVA. In \m{Error in Count}, \f{text} significantly yielded lower error versus \f{opacity} (\p{.0001}, \dvalue{1.74}), \f{size} (\p{.0001}, \dvalue{0.77}), and \f{color} (\p{.05}, \dvalue{0.42}), while \f{size} yielded lower error than \f{opacity} (\p{.0001}, \dvalue{1.46}) and \f{color} (\p{.05}, \dvalue{0.46}). \f{opacity} was rated the worst, even compared to \f{color} (\p{.0001}, \dvalue{1.58}). Similarly, in \m{Error in Least}, \f{text} significantly yielded lower error versus \f{opacity} (\p{.0001}, \dvalue{1.87}), \f{size} (\p{.05}, \dvalue{0.39}), and \f{color} (\p{.0001}, \dvalue{0.98}), while \f{size} yielded lower error than \f{opacity} (\p{.0001}, \dvalue{1.17}) and \f{color} (\p{.0001}, \dvalue{0.74}). \f{opacity} was rated the worst again, even compared to \f{color} (\p{.0005}, \dvalue{0.69}).}

Conversely, in \m{Target Visibility} ({\rmanova{1.89}{73.63}{28.21}{.0001}{0.42}}), participants found \f{opacity} superior, rating it clearer than both \f{text} ({\p{.0001}, \dvalue{0.93}}), \f{size} ({\p{.0001}, \dvalue{1.02}}){, and \f{color} (\p{.05}, \dvalue{0.37})}, while \f{color} having higher ratings than \f{size} ({\p{.0001}, \dvalue{1.01}}) {and \f{text} (\p{.0001}, \dvalue{0.84})}. Combined with the results of the above other measures, these results indicate trade-offs between occlusion and uncertainty recognition when applying signifiers, with those that more clearly conveying uncertainty yielding lower target visibility.

In sum, \f{text} and \f{size} emerged as the most effective signifiers in terms of overall \m{Preference}, \m{Intuitiveness / logic}, \m{Confidence}, \m{Mental Ease}, and error-related measures. Several participants mentioned that for judging uncertainty level and performing clicking tasks, they are \pquote{clear}{p5}, \pquote{easier to compare}{p10}. Many mentioned they would prefer the percentage text design for comparing levels~\textit{[p2-4, 20-25, 28, 36, 39]}; however, participants also mention their concern over \f{text}'s occlusion \pquote{I found the percentages easy to understand but distracting.}{p30}

Even though \f{opacity} and \f{color} are shown as disadvantaged for their limited ability to show visual changes, often associated with lower clarity. However, participants sometimes still prefer them over text because they occluded the targets less. For example, \pquote{Percentages [text] required less effort, but I did not like them as much. My favorite visualization was the one that changed colors.}{p36}


\section{Discussion}
\subsection{Design Recommendations}
{In this section, we distill insights from our survey and experiments, relate them to previous AR studies with uncertainty visualizations, and discuss their implications for selecting archetypes, signifiers, scenes, and their effective synergies.}

\noindent\textit{\textbf{Target visibility \vs{} pointer identifiability.}}
In our findings, we observed divergence between different \UP{}s designs' target visibility and identifiability. Designers should choose pointer types and signifiers based on the cost of a missed coarse selection and the visibility requirement over the targets.

In terms of pointer archetypes, ~{although our surveyed prior literature shows that \f{internal} and \f{fill} pointers are the most common (65\%), our results show they perform poorly when target occlusion is considered}. The two archetypes only outperform the others in scenes with distant targets, where noticeability is more important (shown by clustering in Figure~\ref{fig:uniform_clustering}). Therefore, they are more suitable for input tasks with a critical mis-selection cost. For instance, in fast-moving scenarios, such as driving, where missing a target is more costly, prioritizing disambiguation via more obtrusive pointers may be reasonable.

\noindent\textit{\textbf{Choosing signifier and archetype based on task needs.}}
In terms of Signifiers, for \f{level} pointers, we observed a similar pattern to that of the pointer archetype. Signifiers like \f{text} and \f{size} are more easily perceived and rated highest for task confidence, preference, and level determination. However, although they excel in most measures, \f{size} and \f{text} occlude targets and are even seen as distracting for some participants. On the other hand, \f{opacity} preserves visibility better but provides a weaker gradation of uncertainty intensity. {The observed preference for \f{size} over \f{opacity} contrasts with \citet{kunze_ardriveUV_2018, kunze_drivingsafety_2017}, who found opacity to be preferred over size in a video-based AR study of driving uncertainty. We attribute this difference to task and visualization scale: Kunze et al. applied larger overlays spanning broad roadway regions, where gradations in \f{opacity} are more salient. In contrast, \UP{} visualizations are confined to smaller, pointer-based regions, which likely reduces the perceptual effectiveness of opacity changes.}
Likewise, the \f{boundary} pointer, though rated lowest for level clarity, offered better performance in visibility-sensitive tasks where a layer of visual identity cues is sufficient. For example, in AR-assisted visual search tasks, the ideal signifier should minimize occlusion of the candidate target without overdistraction, allowing users to focus efficiently on recognizing the object; our evidence shows that \f{opacity} or \f{boundary} would suffice.


\noindent\textit{\textbf{Use of pointer archetype depends on conveyed uncertainty information.}} Across the three forms of \UP{}s with different levels of uncertainty information, \f{certain}, \f{identity}, and \f{level}, preferences for archetypes such as \f{boundary} were not consistent.
For instance, the \f{boundary} pointer was the most preferred archetype when visualizing a single, certain target across all four scenes. However, its ranking dropped when used as an \f{identity} pointer in multi-target scenarios, especially under dense or near-target conditions, where separation among targets became harder to distinguish. As a \f{level} pointer, it was among the least preferred, likely because its limited visual area for encoding graded intensity made it less effective at conveying confidence.

\noindent\textit{\textbf{Uniform pointer for accurate coarse selection.}}
We initially included the \f{none}-signifier pointer as a baseline within the \f{identity} pointer category. Although \f{color} and \f{text}-based identity pointers can support direct disambiguation, the \f{none} condition produced significantly better counting accuracy. This suggests that for reliable coarse selection under high input uncertainty, a uniform visualization without identity changes may be more suitable.


\subsection{Additional Example Use}
Besides on-the-go target selection and query over distant or densely-packed objects, as showcased in our teaser figure and introduction, we provide additional example uses of \UP{}s in AR.
\begin{figure*} [t!]
    \centering
    \includegraphics[width=0.7\linewidth]{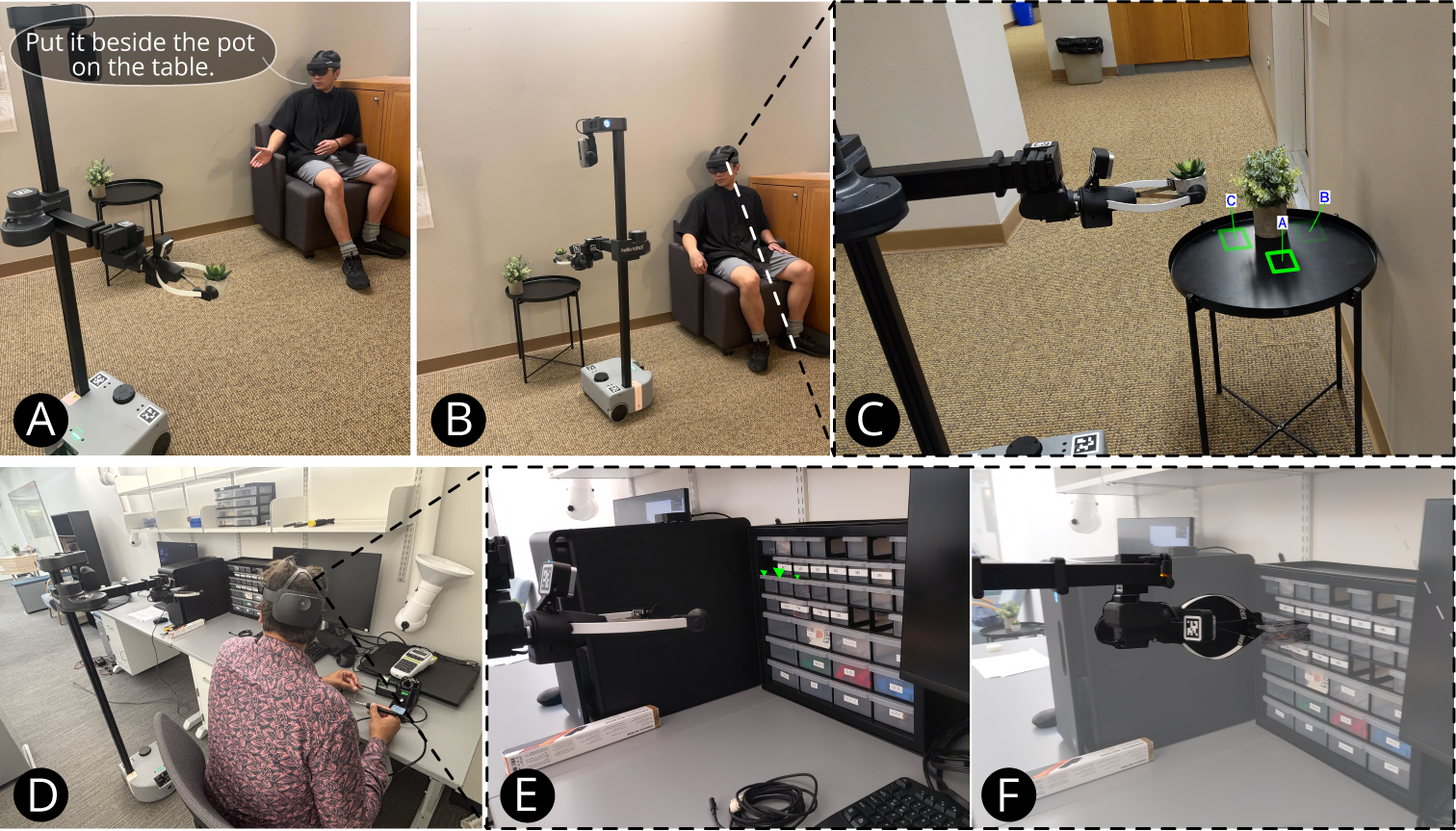}
    
    \caption{Additional example use cases of Uncertain Pointers in AR-HRI.
(A–C) A user tells the robot to put the plant beside the pot. The system detects multiple valid placements due to linguistic ambiguity in ``beside the pot'' and uses \f{boundary} pointers with \f{text} labels to show them, allowing fast user disambiguation.
(D–F) A user with full hands uses gaze as an auxiliary input to ask the robot to fetch tools from a drawer. The system displays \f{level} pointers before action, and the largest pointer points to the wrong one. The user notices the mismatch via this feedforward and adjusts their gaze to clarify intent.}
    \label{fig:example_use}
    \Description{Additional example use cases of Uncertain Pointers in AR-HRI.
(A–C) A user tells the robot to put the plant beside the pot. The system detects multiple valid placements due to linguistic ambiguity in ``beside the pot'' and uses \f{boundary} pointers with \f{text} labels to show them, allowing fast user disambiguation.
(D–F) A user with full hands uses gaze as an auxiliary input to ask the robot to fetch tools from a drawer. The system displays \f{level} pointers before action, and the largest pointer points to the wrong one. The user notices the mismatch via this feedforward and adjusts their gaze to clarify intent.}
\end{figure*}



\noindent\textit{\textbf{Facilitate Communication in Human Robot Interaction.}} Ambiguity in object retrieval or target location is a common challenge in robot perception and command understanding~\cite{pramanick_doro_2022} for human-robot interaction (HRI).
By augmenting existing HRI disambiguation techniques, \UP{}s can serve as a visual communication layer for ambiguity-aware robots, helping them convey uncertainty back to the user. For instance, robots could use \UP{}s to visually present multiple candidate locations when unsure (Figure~\ref{fig:example_use}), allowing users to resolve the ambiguity via additional input. Similarly, \f{level} pointers could signal the system's confidence before the robot commits to costly or irreversible actions. This not only enhances transparency but also fosters more efficient and collaborative interaction between humans and robots.
\looseness=-1

\noindent\textit{\textbf{Composability Beyond the Presented Pointer Space.}} While our evaluation separates \f{level} from \f{identity} and \f{certain} pointers, \UP{}s is not limited to these categories in isolation. In practice, these strategies can be combined when the scenario or target complexity calls for it. For example, as shown in Figure~\ref{fig:example_use}C, a \f{boundary}-\f{opacity} pointer (encoding system confidence) can be augmented with text labels to support disambiguation. This combined approach allows users to both interpret confidence distribution and quickly disambiguate individual targets, enhancing clarity and interaction efficiency. Another example is using a \f{certain} pointer for coarse selection and, upon initial input, displaying candidate target sets with \f{level} or \f{identity} pointers.
\looseness=-1

{

\section{Limitations and Future Work}
\subsection{Video-Based Online Study}
    Our work is the first to identify and systematically explore how uncertainty visualization can be incorporated into situated, object-based pointer design. Our experiments focused on the perception and recognition of different \UP{} designs; however, because we used a screen-based video setup, our study shares several limitations common to prior video-based AR uncertainty visualization research~\cite{kunze_ardriveUV_2018, kunze_drivingsafety_2017}. Below, we outline factors that may cause discrepancies when translating our findings to wearable AR use. Specifically, we consider differences in motion, field of view, lighting, and depth cues in wearable stereoscopic displays.

\noindent\textit{\textbf{Motion.}}
    We simulated walking-in-place motion by applying 2 Hz 6-DoF noise to the camera path~\cite{hirasaki1999effects}. In real use, users may experience translation and can move their bodies more freely, which may reduce overall pointer legibility. We expect the general trends and tradeoffs among pointer and archetype designs to hold. However, for cases with extreme motion, designers might prioritize pointer legibility over real-world visibility, in which case \f{fill} might be a more suitable option despite occluding details of real-world targets. \looseness=-1


\noindent\textit{\textbf{Field of View.}} 
    Lightweight AR devices often have displays with a limited field of view (FoV). Our experiments assume that all candidate targets are within the user's sight and the display's FoV. This assumption aligns well with gaze- and pointing-based selection, where users' attention is likely centered in the FoV. However, for other modalities (e.g., voice queries), target candidates may be far apart and outside the display's FoV, or even out of users' sight entirely. Future work may explore out-of-view cues~\cite{quinn_outofview_2024} to support disambiguation for off-screen targets.


\noindent\textit{\textbf{Lighting.}}
In terms of lighting, our results generalize to virtual reality and passthrough AR experiences. However, with optical see-through AR glasses, users encounter variable lighting conditions. A survey of commercial AR glasses (Section 5.1) revealed an opacity range of 80–100\% for most devices. Accordingly, we capped opacity at 90\% in Study 2. Bright ambient lighting and reflections in the real world reduce contrast and perceived brightness of \UP{}s, worsening the visibility of \f{color} and \f{opacity} signifiers. This will likely make the advantages of \f{text} and \f{size} we observed in Study 2 even more pronounced relative to \f{color} and \f{opacity} signifiers, for communicating uncertainty levels. \looseness=-1 


\noindent\textit{\textbf{Depth Cues.}} 
The apparent size of \UP{}s changes linearly with viewing distance~\cite{holway1941determinants}. To account for this, with our size signifier, we ensure the relative size (\ie{} pixel) is large and clearly visible on our two scenes with sparse targets, utilizing a power function~\cite{stevens_powerlaw_1970}. This approach should generalize to 3D AR. However, when targets span a wide depth range, an even larger step size between levels may be needed to compensate for depth's influence on perceived size.\looseness=-1



\subsection{Visualization Design}}
Our study focused on disambiguation among 3–5 candidate targets, reflecting the ideal noticeable visual changes ($\sim$20\%) of most signifiers ~\cite{oh_size_2020, saturation_threshold, bartram_transparencymin_2011}. However, this does not imply that \UP{}s is only restricted to small candidate sets. For larger sets, \UP{}s can still be used. For instance, using a \f{none} identity pointer to indicate all candidate targets, and applying an additional signifier (\eg{} \f{level} or \f{size}) to emphasize the most probable one, users can still disambiguate effectively through pointing-based input. This approach trades off full uncertainty transparency, but future work can consider multi-level hierarchies by clustering targets and applying leveled variations in signifiers to each cluster or applying more granular and continuous change (\ie{} more than 5 levels) to the entire candidate set.
Additionally, our choice of color (\ie{} green) is based on its preattentive properties and on prior AR work examining contrast against typical backgrounds~\cite{merenda_colorandbg_2016, gabbard_colorblending_2-22}. Future work could investigate adaptive color adjustments to mitigate reduced visibility in specific environments (\eg{} forested scenes) by using alternative preattentive color choices.

{
\subsection{Coverage and Completeness}
\noindent\textit{\textbf{Systematic Review.}} 
Our survey aimed to explore the pointer space of \UP{} by identifying visual signifiers and pointer archetypes commonly used in prior work. A limitation of this approach lies in our chosen venues and query terms; an exhaustive categorization would require surveying an even broader range of publication venues, such as the IEEE/RSJ International Conference on Intelligent Robots and Systems (IROS) to consider uncertain scenarios beyond our current focus, including multi-robot specification (\eg{} \cite{chacon_multirobot_2020}) or IEEE Pacific Visualization Symposium (PacificVis) to incoporate even more uncertainty visualization designs (\eg{} ~\cite{jena}).

\noindent\textit{\textbf{Pointer Space.}}
The presented pointer space is not exhaustive. Although we have coverage of 82\% of the signifiers from the prior work we surveyed, not all signifiers can be described using our current pointer space (\eg{} texture or orientation).
Additionally, future work could explore specific grammars~\cite{wilkinson2011grammar} tailored to object-pointing design with uncertainty information to improve \UP{}s' generative power~\cite{michel_interactionnotinterfaces_2004, michel_instrumentalinteraction_2000}.
}




{
\subsection{\UP{}s with future VLM.}
With recent advances in VLMs, they will likely be integrated into more AR glasses to help resolve uncertain queries. While a stronger reasoning capability may reduce the need for \UP{}s in some cases (\eg{} Figure~\ref{fig:example_use}D-F, where a VLM could infer user task and select the right tool they need), other uncertainty sources remain difficult to eliminate, such as linguistic ambiguity~\cite{wang2025resolving} or cases requiring knowledge of a user’s inherent history or intent (\eg{} which traffic sign a user has not learned in Figure~\ref{fig:teaser}A-B). In these situations, \UP{}s are always beneficial.

As VLMs become better at interpreting complex actions (\eg{} \rev{through demonstration by direct manipulation of images}~\cite{VLMROBOT}), future work on \UP{} could also extend beyond 3D object selection to preview affordances and possible actions in physical or virtual environments with uncertainty cues (\eg{} in combination with ~\cite{VRFF} or ~\cite{Wang2026XOOHRI}). This would provide feedforward support for even more complex task and preview action possibilities.
}


\section{Conclusion}    
We investigated situated feedforward visualizations in AR that annotate multiple real-world objects to represent system uncertainty and support disambiguation by combining pointer designs with existing visualization cues. Through a systematic literature review and two online studies, we analyzed 25 visualization designs across spatial variables. Our findings reveal key trade-offs in how different pointer placements and visual encodings affect user confidence, target visibility, mental load, and more. These insights provide novel insights for designing future ambiguity-aware AR systems for ubiquitous, everyday use.


\begin{acks}
We thank Prof. Manoel Horta Ribeiro for his help with study analysis and are grateful to our thoughtful and insightful reviewers. This work was made possible by the Princeton SEAS Seed Fund.
\end{acks}

\bibliographystyle{ACM-Reference-Format}
\bibliography{_references.bib}


\appendix
\clearpage
\onecolumn
\section{Experiment Interface Supplementary}
\label{sec:appendix-interface}
\begin{figure*} [ht!]
        \centering
        \includegraphics[width=0.6\linewidth]{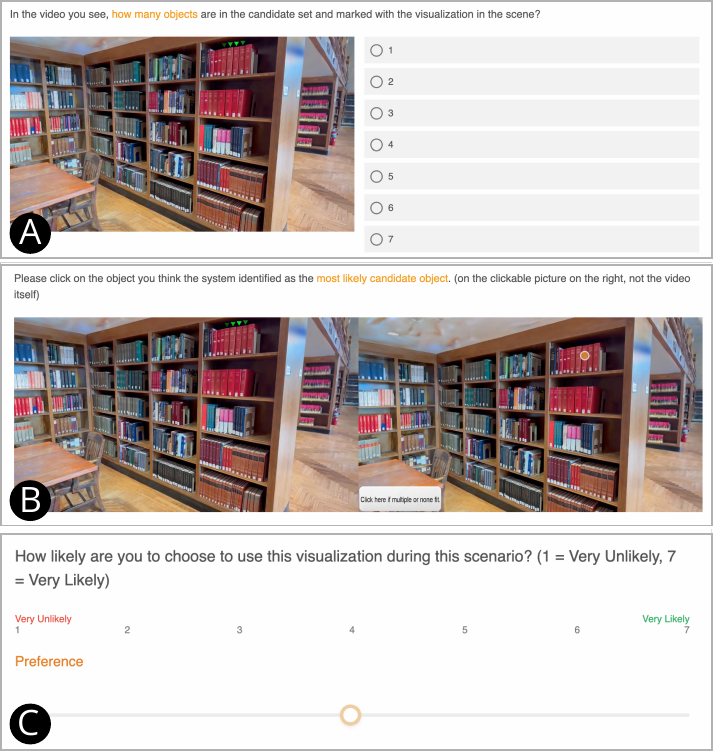}
        \caption{Study interfaces for collecting user response, including the ones for (A) Counting task (Study 1 \& 2), (B) Clicking task (Study 2 only), and (C) Subjective ratings (Study 1 \& 2).}
        \label{fig:study-apparatus}
        \Description{}{}
    \end{figure*}
\newpage
\section{Experiment 1 Statistic Supplementary (\f{archetype} x \f{scene})}
Meandiff = group1 - group2
\label{sec:appendix-experiment1}

\begin{table*}[h]
    \caption{t-test (Holm-Bonferroni-corrected) Result for Confidence}
\begin{tabular}{ccccccc}
\hline
\multicolumn{6}{c}{Confidence}                                                                                                                                                                                                                                                             \\ \hline
\rowcolor[HTML]{FFFFFF} 
{\color[HTML]{434343} group\_1}      & {\color[HTML]{434343} sign}           & {\color[HTML]{434343} group\_2}     & {\color[HTML]{434343} mean\_diff\_(1\_minus\_2)} & {\color[HTML]{434343} cohen\_dz} & {\color[HTML]{434343} p\_holm}\\
\rowcolor[HTML]{F6F8F9} 
{\color[HTML]{434343} Bound x Dense} & {\color[HTML]{434343} \textless{}}    & {\color[HTML]{434343} Bound x Near} & {\color[HTML]{434343} -0.4758}                   & {\color[HTML]{434343} -0.566}    & {\color[HTML]{434343} 0.0058}                  \\
\rowcolor[HTML]{FFFFFF} 
{\color[HTML]{434343} Bound x Dense} & {\color[HTML]{434343} \textless{}}    & {\color[HTML]{434343} Ex x Dense}   & {\color[HTML]{434343} -0.4308}                   & {\color[HTML]{434343} -0.572}    & {\color[HTML]{434343} 0.0049}                  \\
\rowcolor[HTML]{F6F8F9} 
{\color[HTML]{434343} Bound x Far}   & {\color[HTML]{434343} \textgreater{}} & {\color[HTML]{434343} Fill x Dense} & {\color[HTML]{434343} 0.5567}                    & {\color[HTML]{434343} 0.537}     & {\color[HTML]{434343} 0.0122}                  \\
\rowcolor[HTML]{FFFFFF} 
{\color[HTML]{434343} Bound x Near}  & {\color[HTML]{434343} \textgreater{}} & {\color[HTML]{434343} Fill x Dense} & {\color[HTML]{434343} 0.6575}                    & {\color[HTML]{434343} 0.651}     & {\color[HTML]{434343} 0.0006}                  \\
\rowcolor[HTML]{F6F8F9} 
{\color[HTML]{434343} Ex x Dense}    & {\color[HTML]{434343} \textgreater{}} & {\color[HTML]{434343} Fill x Dense} & {\color[HTML]{434343} 0.6125}                    & {\color[HTML]{434343} 0.664}     & {\color[HTML]{434343} 0.0004}                  \\
\bottomrule
\end{tabular}
\end{table*}


\begin{table*}[ht]
    \caption{t-test (Holm-Bonferroni-corrected) Result for Duration}
\begin{tabular}{ccccccc}
\hline
\multicolumn{6}{c}{Duration}                                                                                                                                                                                                                                                       \\ \hline
\rowcolor[HTML]{FFFFFF} 
{\color[HTML]{434343} group\_1}  & {\color[HTML]{434343} sign}        & {\color[HTML]{434343} group\_2}    & {\color[HTML]{434343} mean\_diff\_(1\_minus\_2)} & {\color[HTML]{434343} cohen\_dz} & {\color[HTML]{434343} p\_holm}\\
\rowcolor[HTML]{FFFFFF} 
{\color[HTML]{434343} In x Near} & {\color[HTML]{434343} \textless{}} & {\color[HTML]{434343} In x Sparse} & {\color[HTML]{434343} -0.9363}                   & {\color[HTML]{434343} -0.502}    & {\color[HTML]{434343} 0.0346}                 \\\bottomrule
\end{tabular}
\end{table*}

\begin{table*}[ht]
    \caption{t-test (Holm-Bonferroni-corrected) Result for Mental Ease}
\begin{tabular}{ccccccc}
\hline
\multicolumn{6}{c}{Mental Ease}                                                                                                                                                                                                                                                               \\ \hline
\rowcolor[HTML]{FFFFFF} 
{\color[HTML]{434343} group\_1}       & {\color[HTML]{434343} sign}           & {\color[HTML]{434343} group\_2}       & {\color[HTML]{434343} mean\_diff\_(1\_minus\_2)} & {\color[HTML]{434343} cohen\_dz} & {\color[HTML]{434343} p\_holm}\\
\rowcolor[HTML]{FFFFFF} 
{\color[HTML]{434343} Bound x Dense}  & {\color[HTML]{434343} \textless{}}    & {\color[HTML]{434343} Bound x Far}    & {\color[HTML]{434343} -0.9475}                   & {\color[HTML]{434343} -0.685}    & {\color[HTML]{434343} 0.0002}                  \\
\rowcolor[HTML]{F6F8F9} 
{\color[HTML]{434343} Bound x Dense}  & {\color[HTML]{434343} \textless{}}    & {\color[HTML]{434343} Bound x Near}   & {\color[HTML]{434343} -1.2925}                   & {\color[HTML]{434343} -0.8}      & {\color[HTML]{434343} 0}                       \\
\rowcolor[HTML]{FFFFFF} 
{\color[HTML]{434343} Bound x Dense}  & {\color[HTML]{434343} \textless{}}    & {\color[HTML]{434343} Bound x Sparse} & {\color[HTML]{434343} -0.9442}                   & {\color[HTML]{434343} -0.6}      & {\color[HTML]{434343} 0.002}                   \\
\rowcolor[HTML]{F6F8F9} 
{\color[HTML]{434343} Bound x Dense}  & {\color[HTML]{434343} \textless{}}    & {\color[HTML]{434343} Ex x Dense}     & {\color[HTML]{434343} -0.9825}                   & {\color[HTML]{434343} -0.825}    & {\color[HTML]{434343} 0}                       \\
\rowcolor[HTML]{FFFFFF} 
{\color[HTML]{434343} Bound x Dense}  & {\color[HTML]{434343} \textless{}}    & {\color[HTML]{434343} Ex x Near}      & {\color[HTML]{434343} -1.0703}                   & {\color[HTML]{434343} -0.817}    & {\color[HTML]{434343} 0}                       \\
\bottomrule                 
\end{tabular}
\end{table*}

\begin{table*}[ht]
    \caption{t-test (Holm-Bonferroni-corrected) Result for Target Visibility (1)}
\begin{tabular}{ccccccc}
\hline
\multicolumn{6}{c}{Target Visibility}                                                                                                                                                                                                                                                        \\ \hline
\rowcolor[HTML]{FFFFFF} 
{\color[HTML]{434343} group\_1}       & {\color[HTML]{434343} sign}           & {\color[HTML]{434343} group\_2}      & {\color[HTML]{434343} mean\_diff\_(1\_minus\_2)} & {\color[HTML]{434343} cohen\_dz} & {\color[HTML]{434343} p\_holm}\\
\rowcolor[HTML]{F6F8F9} 
{\color[HTML]{434343} Bound x Dense}  & {\color[HTML]{434343} \textgreater{}} & {\color[HTML]{434343} Fill x Dense}  & {\color[HTML]{434343} 2.8258}                    & {\color[HTML]{434343} 1.423}     & {\color[HTML]{434343} 0}                       \\
\rowcolor[HTML]{FFFFFF} 
{\color[HTML]{434343} Bound x Dense}  & {\color[HTML]{434343} \textgreater{}} & {\color[HTML]{434343} Fill x Far}    & {\color[HTML]{434343} 2.5142}                    & {\color[HTML]{434343} 1.26}      & {\color[HTML]{434343} 0}                       \\
\rowcolor[HTML]{F6F8F9} 
{\color[HTML]{434343} Bound x Dense}  & {\color[HTML]{434343} \textgreater{}} & {\color[HTML]{434343} Fill x Near}   & {\color[HTML]{434343} 2.7217}                    & {\color[HTML]{434343} 1.148}     & {\color[HTML]{434343} 0}                       \\
\rowcolor[HTML]{FFFFFF} 
{\color[HTML]{434343} Bound x Dense}  & {\color[HTML]{434343} \textgreater{}} & {\color[HTML]{434343} Fill x Sparse} & {\color[HTML]{434343} 2.6}                       & {\color[HTML]{434343} 1.22}      & {\color[HTML]{434343} 0}                       \\
\rowcolor[HTML]{F6F8F9} 
{\color[HTML]{434343} Bound x Dense}  & {\color[HTML]{434343} \textgreater{}} & {\color[HTML]{434343} In x Dense}    & {\color[HTML]{434343} 0.9131}                    & {\color[HTML]{434343} 0.628}     & {\color[HTML]{434343} 0.0005}                  \\
\bottomrule        
\end{tabular}
\end{table*}

\begin{table*}[ht]
    \caption{t-test (Holm-Bonferroni-corrected) Result for Target Visibility (2)}
\begin{tabular}{ccccccc}
\hline
\multicolumn{6}{c}{Target Visibility}                                                                                                                                                                                                                                                        \\ \hline
\rowcolor[HTML]{FFFFFF} 
{\color[HTML]{434343} group\_1}       & {\color[HTML]{434343} sign}           & {\color[HTML]{434343} group\_2}      & {\color[HTML]{434343} mean\_diff\_(1\_minus\_2)} & {\color[HTML]{434343} cohen\_dz} & {\color[HTML]{434343} p\_holm}\\
\rowcolor[HTML]{F6F8F9} 
{\color[HTML]{434343} Ex x Far}       & {\color[HTML]{434343} \textgreater{}} & {\color[HTML]{434343} Fill x Near}   & {\color[HTML]{434343} 2.3681}                    & {\color[HTML]{434343} 1.136}     & {\color[HTML]{434343} 0}                       \\
\rowcolor[HTML]{FFFFFF} 
{\color[HTML]{434343} Ex x Far}       & {\color[HTML]{434343} \textgreater{}} & {\color[HTML]{434343} Fill x Sparse} & {\color[HTML]{434343} 2.2464}                    & {\color[HTML]{434343} 1.172}     & {\color[HTML]{434343} 0}                       \\
\rowcolor[HTML]{F6F8F9} 
{\color[HTML]{434343} Ex x Far}       & {\color[HTML]{434343} \textgreater{}} & {\color[HTML]{434343} In x Near}     & {\color[HTML]{434343} 2.2256}                    & {\color[HTML]{434343} 1.195}     & {\color[HTML]{434343} 0}                       \\
\rowcolor[HTML]{FFFFFF} 
{\color[HTML]{434343} Ex x Near}      & {\color[HTML]{434343} \textgreater{}} & {\color[HTML]{434343} Ex x Sparse}   & {\color[HTML]{434343} 0.3917}                    & {\color[HTML]{434343} 0.445}     & {\color[HTML]{434343} 0.0473}                  \\
\rowcolor[HTML]{F6F8F9} 
{\color[HTML]{434343} Ex x Near}      & {\color[HTML]{434343} \textgreater{}} & {\color[HTML]{434343} Fill x Dense}  & {\color[HTML]{434343} 3.2256}                    & {\color[HTML]{434343} 1.742}     & {\color[HTML]{434343} 0}                       \\
\bottomrule        
\end{tabular}
\end{table*}

\begin{table*}[ht]
    \caption{t-test (Holm-Bonferroni-corrected) Result for Preference (1)}
\begin{tabular}{ccccccc}
\hline
\multicolumn{6}{c}{Preference}                                                                                                                                                                                                                                                                \\ \hline
\rowcolor[HTML]{FFFFFF} 
{\color[HTML]{434343} group\_1}       & {\color[HTML]{434343} sign}           & {\color[HTML]{434343} group\_2}       & {\color[HTML]{434343} mean\_diff\_(1\_minus\_2)} & {\color[HTML]{434343} cohen\_dz} & {\color[HTML]{434343} p\_holm}\\
\rowcolor[HTML]{FFFFFF} 
{\color[HTML]{434343} Bound x Dense}  & {\color[HTML]{434343} \textless{}}    & {\color[HTML]{434343} Bound x Far}    & {\color[HTML]{434343} -1.0292}                   & {\color[HTML]{434343} -0.603}    & {\color[HTML]{434343} 0.001}                   \\
\rowcolor[HTML]{F6F8F9} 
{\color[HTML]{434343} Bound x Dense}  & {\color[HTML]{434343} \textless{}}    & {\color[HTML]{434343} Bound x Near}   & {\color[HTML]{434343} -1.5317}                   & {\color[HTML]{434343} -1.016}    & {\color[HTML]{434343} 0}                       \\
\rowcolor[HTML]{FFFFFF} 
{\color[HTML]{434343} Bound x Dense}  & {\color[HTML]{434343} \textless{}}    & {\color[HTML]{434343} Bound x Sparse} & {\color[HTML]{434343} -1.1667}                   & {\color[HTML]{434343} -0.68}     & {\color[HTML]{434343} 0.0001}                  \\
\rowcolor[HTML]{F6F8F9} 
{\color[HTML]{434343} Bound x Dense}  & {\color[HTML]{434343} \textless{}}    & {\color[HTML]{434343} Ex x Dense}     & {\color[HTML]{434343} -1.0047}                   & {\color[HTML]{434343} -0.676}    & {\color[HTML]{434343} 0.0001}                  \\
\rowcolor[HTML]{FFFFFF} 
{\color[HTML]{434343} Bound x Dense}  & {\color[HTML]{434343} \textless{}}    & {\color[HTML]{434343} Ex x Near}      & {\color[HTML]{434343} -1.0481}                   & {\color[HTML]{434343} -0.742}    & {\color[HTML]{434343} 0}                       \\
\bottomrule              
\end{tabular}
\end{table*}

\begin{table*}[ht]
    \caption{t-test (Holm-Bonferroni-corrected) Result for Preference (2)}
\begin{tabular}{ccccccc}
\hline
\multicolumn{6}{c}{Preference}                                           \\ \hline
\rowcolor[HTML]{FFFFFF} 
{\color[HTML]{434343} group\_1}       & {\color[HTML]{434343} sign}           & {\color[HTML]{434343} group\_2}       & {\color[HTML]{434343} mean\_diff\_(1\_minus\_2)} & {\color[HTML]{434343} cohen\_dz} & {\color[HTML]{434343} p\_holm}\\
\rowcolor[HTML]{FFFFFF} 
{\color[HTML]{434343} Ex x Dense}     & {\color[HTML]{434343} \textgreater{}} & {\color[HTML]{434343} Fill x Sparse}  & {\color[HTML]{434343} 1.8714}                    & {\color[HTML]{434343} 1.038}     & {\color[HTML]{434343} 0}                       \\
\rowcolor[HTML]{F6F8F9} 
{\color[HTML]{434343} Ex x Dense}     & {\color[HTML]{434343} \textgreater{}} & {\color[HTML]{434343} In x Dense}     & {\color[HTML]{434343} 1.0589}                    & {\color[HTML]{434343} 0.89}      & {\color[HTML]{434343} 0}                       \\
\rowcolor[HTML]{FFFFFF} 
{\color[HTML]{434343} Ex x Dense}     & {\color[HTML]{434343} \textgreater{}} & {\color[HTML]{434343} In x Far}       & {\color[HTML]{434343} 0.8667}                    & {\color[HTML]{434343} 0.706}     & {\color[HTML]{434343} 0.0001}                  \\
\rowcolor[HTML]{F6F8F9} 
{\color[HTML]{434343} Ex x Dense}     & {\color[HTML]{434343} \textgreater{}} & {\color[HTML]{434343} In x Near}      & {\color[HTML]{434343} 2.1861}                    & {\color[HTML]{434343} 1.455}     & {\color[HTML]{434343} 0}                       \\
\rowcolor[HTML]{FFFFFF} 
{\color[HTML]{434343} Ex x Dense}     & {\color[HTML]{434343} \textgreater{}} & {\color[HTML]{434343} In x Sparse}    & {\color[HTML]{434343} 1.215}                     & {\color[HTML]{434343} 0.882}     & {\color[HTML]{434343} 0}                       \\
\bottomrule              
\end{tabular}
\end{table*}

\clearpage
\newpage
\section{Experiment 2 Statistic Supplementary (\f{archetype} x \f{scene})}
Meandiff = group1 - group2
\label{sec:appendix-experiment2}

\begin{table*}[h]
    \caption{t-test (Holm-Bonferroni-corrected) Result for Confidence}
\begin{tabular}{ccccccc}
\hline
\multicolumn{6}{c}{Confidence}                                                                                                                                                                                                                                                           \\ \hline
\rowcolor[HTML]{F6F8F9} 
{\color[HTML]{434343} group\_1}       & {\color[HTML]{434343} sign}        & {\color[HTML]{434343} group\_2}     & {\color[HTML]{434343} mean\_diff\_(1\_minus\_2)} & {\color[HTML]{434343} cohen\_dz} & {\color[HTML]{434343} p\_holm}\\
\rowcolor[HTML]{F6F8F9} 
{\color[HTML]{434343} Bound x Dense}  & {\color[HTML]{434343} \textless{}} & {\color[HTML]{434343} Ex x Dense}   & {\color[HTML]{434343} -0.695}                    & {\color[HTML]{434343} -0.61}     & {\color[HTML]{434343} 0.0155}                  \\
\rowcolor[HTML]{FFFFFF} 
{\color[HTML]{434343} Bound x Dense}  & {\color[HTML]{434343} \textless{}} & {\color[HTML]{434343} Ex x Near}    & {\color[HTML]{434343} -0.6931}                   & {\color[HTML]{434343} -0.594}    & {\color[HTML]{434343} 0.0195}                  \\
\rowcolor[HTML]{F6F8F9} 
{\color[HTML]{434343} Bound x Far}    & {\color[HTML]{434343} \textless{}} & {\color[HTML]{434343} Bound x Near} & {\color[HTML]{434343} -0.6583}                   & {\color[HTML]{434343} -0.608}    & {\color[HTML]{434343} 0.0155}                  \\
\rowcolor[HTML]{FFFFFF} 
{\color[HTML]{434343} Bound x Far}    & {\color[HTML]{434343} \textless{}} & {\color[HTML]{434343} Ex x Dense}   & {\color[HTML]{434343} -0.9792}                   & {\color[HTML]{434343} -0.974}    & {\color[HTML]{434343} 0}                       \\
\rowcolor[HTML]{F6F8F9} 
{\color[HTML]{434343} Bound x Far}    & {\color[HTML]{434343} \textless{}} & {\color[HTML]{434343} Ex x Far}     & {\color[HTML]{434343} -0.7317}                   & {\color[HTML]{434343} -0.757}    & {\color[HTML]{434343} 0.001}                   \\
\bottomrule                 
\end{tabular}
\end{table*}

\begin{table*}[h]
    \caption{t-test (Holm-Bonferroni-corrected) Result for Intuitiveness / Logic}
\begin{tabular}{ccccccc}
\hline
\multicolumn{6}{c}{Intuitiveness / Logic}                                                                                                                                                                                                                                               \\ \hline
\rowcolor[HTML]{F6F8F9} 
{\color[HTML]{434343} group\_1}       & {\color[HTML]{434343} sign}        & {\color[HTML]{434343} group\_2}    & {\color[HTML]{434343} mean\_diff\_(1\_minus\_2)} & {\color[HTML]{434343} cohen\_dz} & {\color[HTML]{434343} p\_holm}\\
\rowcolor[HTML]{F6F8F9} 
{\color[HTML]{434343} Bound x Dense}  & {\color[HTML]{434343} \textless{}} & {\color[HTML]{434343} Ex x Dense}  & {\color[HTML]{434343} -0.753}                    & {\color[HTML]{434343} -0.721}    & {\color[HTML]{434343} 0.0027}                  \\
\rowcolor[HTML]{FFFFFF} 
{\color[HTML]{434343} Bound x Dense}  & {\color[HTML]{434343} \textless{}} & {\color[HTML]{434343} Ex x Near}   & {\color[HTML]{434343} -0.6145}                   & {\color[HTML]{434343} -0.599}    & {\color[HTML]{434343} 0.0222}                  \\
\rowcolor[HTML]{F6F8F9} 
{\color[HTML]{434343} Bound x Far}    & {\color[HTML]{434343} \textless{}} & {\color[HTML]{434343} Ex x Dense}  & {\color[HTML]{434343} -0.9855}                   & {\color[HTML]{434343} -0.791}    & {\color[HTML]{434343} 0.0007}                  \\
\rowcolor[HTML]{FFFFFF} 
{\color[HTML]{434343} Bound x Far}    & {\color[HTML]{434343} \textless{}} & {\color[HTML]{434343} Ex x Far}    & {\color[HTML]{434343} -0.7312}                   & {\color[HTML]{434343} -0.707}    & {\color[HTML]{434343} 0.0033}                  \\
\rowcolor[HTML]{F6F8F9} 
{\color[HTML]{434343} Bound x Far}    & {\color[HTML]{434343} \textless{}} & {\color[HTML]{434343} Ex x Near}   & {\color[HTML]{434343} -0.847}                    & {\color[HTML]{434343} -0.688}    & {\color[HTML]{434343} 0.0047}                  \\
\bottomrule                  
\end{tabular}
\end{table*}

\begin{table*}[h]
    \caption{t-test (Holm-Bonferroni-corrected) Result for Mental Ease}
\begin{tabular}{ccccccc}
\hline
\multicolumn{6}{c}{Mental Ease}                                                                                                                                                                                                                                                        \\ \hline
\rowcolor[HTML]{F6F8F9} 
{\color[HTML]{434343} group\_1}      & {\color[HTML]{434343} sign}        & {\color[HTML]{434343} group\_2}    & {\color[HTML]{434343} mean\_diff\_(1\_minus\_2)} & {\color[HTML]{434343} cohen\_dz} & {\color[HTML]{434343} p\_holm}\\
\rowcolor[HTML]{F6F8F9} 
{\color[HTML]{434343} Bound x Dense} & {\color[HTML]{434343} \textless{}} & {\color[HTML]{434343} Ex x Dense}  & {\color[HTML]{434343} -0.5904}                   & {\color[HTML]{434343} -0.553}    & {\color[HTML]{434343} 0.048}                   \\
\rowcolor[HTML]{FFFFFF} 
{\color[HTML]{434343} Bound x Far}   & {\color[HTML]{434343} \textless{}} & {\color[HTML]{434343} Ex x Dense}  & {\color[HTML]{434343} -0.9471}                   & {\color[HTML]{434343} -0.834}    & {\color[HTML]{434343} 0.0002}                  \\
\rowcolor[HTML]{F6F8F9} 
{\color[HTML]{434343} Bound x Far}   & {\color[HTML]{434343} \textless{}} & {\color[HTML]{434343} Ex x Far}    & {\color[HTML]{434343} -0.5733}                   & {\color[HTML]{434343} -0.554}    & {\color[HTML]{434343} 0.048}                   \\
\rowcolor[HTML]{FFFFFF} 
{\color[HTML]{434343} Bound x Far}   & {\color[HTML]{434343} \textless{}} & {\color[HTML]{434343} Ex x Near}   & {\color[HTML]{434343} -0.9058}                   & {\color[HTML]{434343} -0.792}    & {\color[HTML]{434343} 0.0005}                  \\
\rowcolor[HTML]{F6F8F9} 
{\color[HTML]{434343} Bound x Far}   & {\color[HTML]{434343} \textless{}} & {\color[HTML]{434343} Ex x Sparse} & {\color[HTML]{434343} -0.7696}                   & {\color[HTML]{434343} -0.636}    & {\color[HTML]{434343} 0.0109}                  \\
\bottomrule                
\end{tabular}
\end{table*}

\begin{table*}[ht]
    \caption{t-test (Holm-Bonferroni-corrected) Result for Target Visibility}
\begin{tabular}{ccccccc}
\hline
\multicolumn{6}{c}{Target Visibility}                                                                                                                                                                                                                                                       \\ \hline
\rowcolor[HTML]{F6F8F9} 
{\color[HTML]{434343} group\_1}       & {\color[HTML]{434343} sign}           & {\color[HTML]{434343} group\_2}     & {\color[HTML]{434343} mean\_diff\_(1\_minus\_2)} & {\color[HTML]{434343} cohen\_dz} & {\color[HTML]{434343} p\_holm}\\
\rowcolor[HTML]{F6F8F9} 
{\color[HTML]{434343} Bound x Dense}  & {\color[HTML]{434343} \textless{}}    & {\color[HTML]{434343} Bound x Near} & {\color[HTML]{434343} -0.6033}                   & {\color[HTML]{434343} -0.538}    & {\color[HTML]{434343} 0.0449}                  \\
\rowcolor[HTML]{FFFFFF} 
{\color[HTML]{434343} Bound x Dense}  & {\color[HTML]{434343} \textless{}}    & {\color[HTML]{434343} Ex x Dense}   & {\color[HTML]{434343} -0.835}                    & {\color[HTML]{434343} -0.992}    & {\color[HTML]{434343} 0}                       \\
\rowcolor[HTML]{F6F8F9} 
{\color[HTML]{434343} Bound x Dense}  & {\color[HTML]{434343} \textless{}}    & {\color[HTML]{434343} Ex x Near}    & {\color[HTML]{434343} -0.8719}                   & {\color[HTML]{434343} -0.711}    & {\color[HTML]{434343} 0.0021}                  \\
\rowcolor[HTML]{FFFFFF} 
{\color[HTML]{434343} Bound x Far}    & {\color[HTML]{434343} \textless{}}    & {\color[HTML]{434343} Ex x Dense}   & {\color[HTML]{434343} -0.8117}                   & {\color[HTML]{434343} -0.78}     & {\color[HTML]{434343} 0.0006}                  \\
\rowcolor[HTML]{F6F8F9} 
{\color[HTML]{434343} Bound x Far}    & {\color[HTML]{434343} \textless{}}    & {\color[HTML]{434343} Ex x Near}    & {\color[HTML]{434343} -0.8485}                   & {\color[HTML]{434343} -0.745}    & {\color[HTML]{434343} 0.0011}                  \\
\bottomrule                
\end{tabular}
\end{table*}

\begin{table*}[ht]
    \caption{t-test (Holm-Bonferroni-corrected) Result for Preference}
\begin{tabular}{ccccccc}
\hline
\multicolumn{6}{c}{Preference}                                                                                                                                                                                                                                                                \\ \hline
\rowcolor[HTML]{F6F8F9} 
{\color[HTML]{434343} group\_1}       & {\color[HTML]{434343} sign}           & {\color[HTML]{434343} group\_2}       & {\color[HTML]{434343} mean\_diff\_(1\_minus\_2)} & {\color[HTML]{434343} cohen\_dz} & {\color[HTML]{434343} p\_holm}\\
\rowcolor[HTML]{F6F8F9} 
{\color[HTML]{434343} Bound x Dense}  & {\color[HTML]{434343} \textless{}}    & {\color[HTML]{434343} Bound x Near}   & {\color[HTML]{434343} -0.84}                     & {\color[HTML]{434343} -0.637}    & {\color[HTML]{434343} 0.0086}                  \\
\rowcolor[HTML]{FFFFFF} 
{\color[HTML]{434343} Bound x Dense}  & {\color[HTML]{434343} \textless{}}    & {\color[HTML]{434343} Ex x Dense}     & {\color[HTML]{434343} -1.2329}                   & {\color[HTML]{434343} -0.971}    & {\color[HTML]{434343} 0}                       \\
\rowcolor[HTML]{F6F8F9} 
{\color[HTML]{434343} Bound x Dense}  & {\color[HTML]{434343} \textless{}}    & {\color[HTML]{434343} Ex x Near}      & {\color[HTML]{434343} -0.906}                    & {\color[HTML]{434343} -0.636}    & {\color[HTML]{434343} 0.0086}                  \\
\rowcolor[HTML]{FFFFFF} 
{\color[HTML]{434343} Bound x Far}    & {\color[HTML]{434343} \textless{}}    & {\color[HTML]{434343} Bound x Near}   & {\color[HTML]{434343} -1.1408}                   & {\color[HTML]{434343} -0.941}    & {\color[HTML]{434343} 0}                       \\
\rowcolor[HTML]{F6F8F9} 
{\color[HTML]{434343} Bound x Far}    & {\color[HTML]{434343} \textless{}}    & {\color[HTML]{434343} Ex x Dense}     & {\color[HTML]{434343} -1.5338}                   & {\color[HTML]{434343} -1.05}     & {\color[HTML]{434343} 0}                       \\
\bottomrule                
\end{tabular}

\end{table*}

\begin{table*}[ht]
    \caption{t-test (Holm-Bonferroni-corrected) Result for Error in Least}
\begin{tabular}{ccccccc}
\hline
\multicolumn{6}{c}{Error in Least}                                                                                                                                                                                                                                                            \\ \hline
\rowcolor[HTML]{F6F8F9} 
{\color[HTML]{434343} group\_1}       & {\color[HTML]{434343} sign}           & {\color[HTML]{434343} group\_2}       & {\color[HTML]{434343} mean\_diff\_(1\_minus\_2)} & {\color[HTML]{434343} cohen\_dz} & {\color[HTML]{434343} p\_holm}\\
\rowcolor[HTML]{FFFFFF} 
{\color[HTML]{434343} Bound x Dense}  & {\color[HTML]{434343} \textgreater{}} & {\color[HTML]{434343} Bound x Sparse} & {\color[HTML]{434343} 0.3782}                    & {\color[HTML]{434343} 0.884}     & {\color[HTML]{434343} 0.0001}                  \\
\rowcolor[HTML]{F6F8F9} 
{\color[HTML]{434343} Bound x Dense}  & {\color[HTML]{434343} \textgreater{}} & {\color[HTML]{434343} Ex x Dense}     & {\color[HTML]{434343} 0.3974}                    & {\color[HTML]{434343} 1.093}     & {\color[HTML]{434343} 0}                       \\
\rowcolor[HTML]{FFFFFF} 
{\color[HTML]{434343} Bound x Dense}  & {\color[HTML]{434343} \textgreater{}} & {\color[HTML]{434343} Ex x Near}      & {\color[HTML]{434343} 0.3066}                    & {\color[HTML]{434343} 0.626}     & {\color[HTML]{434343} 0.0124}                  \\
\rowcolor[HTML]{F6F8F9} 
{\color[HTML]{434343} Bound x Dense}  & {\color[HTML]{434343} \textgreater{}} & {\color[HTML]{434343} Ex x Sparse}    & {\color[HTML]{434343} 0.4231}                    & {\color[HTML]{434343} 1.1}       & {\color[HTML]{434343} 0}                       \\
\rowcolor[HTML]{FFFFFF} 
{\color[HTML]{434343} Bound x Dense}  & {\color[HTML]{434343} \textgreater{}} & {\color[HTML]{434343} In x Sparse}    & {\color[HTML]{434343} 0.3803}                    & {\color[HTML]{434343} 0.815}     & {\color[HTML]{434343} 0.0004}                  \\
\bottomrule               
\end{tabular}

\end{table*}

\begin{table*}[ht]
    \caption{t-test (Holm-Bonferroni-corrected) Result for Error in Count}
\begin{tabular}{ccccccc}
\hline
\multicolumn{6}{c}{Error in Count}                                                                                                                                                                                                                                                            \\ \hline
\rowcolor[HTML]{F6F8F9} 
{\color[HTML]{434343} group\_1}       & {\color[HTML]{434343} sign}           & {\color[HTML]{434343} group\_2}       & {\color[HTML]{434343} mean\_diff\_(1\_minus\_2)} & {\color[HTML]{434343} cohen\_dz} & {\color[HTML]{434343} p\_holm}\\
\rowcolor[HTML]{F6F8F9} 
{\color[HTML]{434343} Bound x Dense}  & {\color[HTML]{434343} \textgreater{}} & {\color[HTML]{434343} Bound x Sparse} & {\color[HTML]{434343} 0.2583}                    & {\color[HTML]{434343} 0.843}     & {\color[HTML]{434343} 0.0002}                  \\
\rowcolor[HTML]{FFFFFF} 
{\color[HTML]{434343} Bound x Dense}  & {\color[HTML]{434343} \textgreater{}} & {\color[HTML]{434343} Ex x Dense}     & {\color[HTML]{434343} 0.375}                     & {\color[HTML]{434343} 1.423}     & {\color[HTML]{434343} 0}                       \\
\rowcolor[HTML]{F6F8F9} 
{\color[HTML]{434343} Bound x Dense}  & {\color[HTML]{434343} \textgreater{}} & {\color[HTML]{434343} Ex x Near}      & {\color[HTML]{434343} 0.2312}                    & {\color[HTML]{434343} 0.857}     & {\color[HTML]{434343} 0.0001}                  \\
\rowcolor[HTML]{FFFFFF} 
{\color[HTML]{434343} Bound x Dense}  & {\color[HTML]{434343} \textgreater{}} & {\color[HTML]{434343} Ex x Sparse}    & {\color[HTML]{434343} 0.2625}                    & {\color[HTML]{434343} 0.847}     & {\color[HTML]{434343} 0.0002}                  \\
\rowcolor[HTML]{F6F8F9} 
{\color[HTML]{434343} Bound x Dense}  & {\color[HTML]{434343} \textgreater{}} & {\color[HTML]{434343} In x Sparse}    & {\color[HTML]{434343} 0.2188}                    & {\color[HTML]{434343} 0.615}     & {\color[HTML]{434343} 0.0126}                  \\
\bottomrule                
\end{tabular}

\end{table*}


\end{document}